\documentclass[twocolumn, twocolappendix, appendixfloats, numberedappendix, apj]{openjournal}

\usepackage{latexsym}
\usepackage{graphicx}
\usepackage{amssymb}
\usepackage{longtable}
\usepackage{epsf}
\usepackage{amsmath}
\usepackage{graphicx}
\usepackage{hyperref}
\hypersetup{colorlinks=true,linkcolor=BlueViolet,citecolor=BlueViolet,filecolor=BlueViolet,urlcolor=BlueViolet}
\usepackage{cleveref}
\usepackage{bm}
\usepackage{lipsum}
\usepackage{enumitem}
\usepackage{newtx, newtxtext, newtxmath}
\usepackage{gensymb}
\usepackage{booktabs}
\usepackage{multirow}
\usepackage{natbib}
\usepackage{fnpct}
\setfnpct{reverse}
\usepackage{float}
\usepackage{xspace}
\usepackage{cleveref}

\usepackage[dvipsnames]{xcolor}

\newcommand{\eg}{\textit{e.g.}\xspace}
\newcommand{\ie}{\textit{i.e.}\xspace}

\newcommand{\picasso}{\texttt{picasso}\xspace}
\newcommand\smaller[2][0.85]{{\scalefont{#1}#2}}
\newcommand{\jax}{\texttt{jax}\xspace}
\newcommand{\HACC}{\smaller{HACC}\xspace}

\newcommand{\CRK}{\smaller{CRK}\xspace}

\newcommand{\rhog}{\rho_{\rm g}}
\newcommand{\Ptot}{P_{\rm tot}}
\newcommand{\Pth}{P_{\rm th}}
\newcommand{\fnt}{f_{\rm nt}}
\newcommand{\Ant}{A_{\rm nt}}
\newcommand{\Bnt}{B_{\rm nt}}
\newcommand{\Cnt}{C_{\rm nt}}
\newcommand{\NR}{{\rm NR}}
\newcommand{\SG}{{\rm SG}}
\newcommand{\vshell}{V_{\rm shell}}
\newcommand{\rhocrit}{\rho_{\rm crit.}}
\newcommand{\thetahalo}{\vartheta_{\rm halo}}
\newcommand{\thetahaloi}{\vartheta_{{\rm halo}, \, i}}
\newcommand{\thetagas}{\vartheta_{\rm gas}}
\newcommand{\thetagasi}{\vartheta_{{\rm gas}, \, i}}
\newcommand{\thetanet}{\vartheta_{\rm nn}}
\newcommand{\yay}{\textcolor{Green}{\textbf{\checkmark}}}
\newcommand{\nay}{\textcolor{BrickRed}{$\times$}}

\newcommand{\refmodeleqs}{eqs.~(\ref{eq:model_rho_P}--\ref{eq:model_pth})\xspace}

\newcommand{\rfnt}{2R_{500c}}
\newcommand{\fullrrange}{$r/R_{500c} \in [0.1, \, 2.0]$}

\defcitealias{2023OJAp....6E..43K}{K23}
\newcommand{\bpbc}{\citetalias{2023OJAp....6E..43K}\xspace}

\begin{document}

\title{The picasso gas model: Painting intracluster gas on gravity-only simulations\vspace{-1.4cm}}
\shorttitle{The \picasso gas model}

\author{F.~K\'eruzor\'e$^{1,\dagger}$}
\author{L.~E.~Bleem$^1$}
\author{N.~Frontiere$^2$}
\author{N.~Krishnan$^1$}
\author{M.~Buehlmann$^2$}
\author{J.D.~Emberson$^2$}
\author{S.~Habib$^{1,2}$}
\author{P.~Larsen$^2$}

\affiliation{$^1$ High Energy Physics Division, Argonne National Laboratory, Lemont, IL 60439, USA}
\affiliation{$^2$ Computational Science Division, Argonne National Laboratory, Lemont, IL 60439, USA}
\thanks{$^{\dagger}$e-mail:fkeruzore@anl.gov}

\shortauthors{K\'eruzor\'e et al.}


\begin{abstract}
    We introduce \picasso, a model designed to predict thermodynamic properties of the intracluster medium based on the properties of halos in gravity-only simulations.
    The predictions result from the combination of an analytical gas model, mapping gas properties to the gravitational potential, and of a machine learning model to predict the model parameters for individual halos based on their scalar properties, such as mass and concentration.
    Once trained, the model can be applied to make predictions for arbitrary potential distributions, allowing its use with flexible inputs such as $N-$body particle distributions or radial profiles.
    We present the model, and train it using pairs of gravity-only and hydrodynamic simulations.
    We show that when trained to learn the mapping from gravity-only to non-radiative hydrodynamic simulations, \picasso can make remarkably accurate and precise predictions of intracluster gas thermodynamics, with percent-level bias and $\sim 20\%$ scatter for $r/R_{500c} \in [0.1, 1]$.
    Training the model on hydrodynamic simulations including sub-resolution physics modeling yields robust predictions as well, albeit with the introduction of a radius-dependent bias and an increase in scatter.
    We further show that the model can be trained to make accurate predictions from very minimal halo information, down to mass and concentration, at the cost of modestly reduced precision.
    \picasso is made publicly available on Github\footnote{\url{https://github.com/fkeruzore/picasso}} as a Python package, which includes trained models that can be used to make predictions easily and efficiently, in a fully auto-differentiable and hardware-accelerated framework.
\end{abstract}


\maketitle

\vspace{1cm}

\twocolumngrid

%


\section{Introduction} \label{sec:intro}

The abundance of dark matter halos in mass across cosmic time is extremely sensitive to the underlying cosmological parameters.
Consequently, the distribution of galaxy clusters---hosted by the most massive dark matter halos---in mass and redshift is a powerful cosmological probe \citep[see, \eg,][for a review]{2011ARA&A..49..409A}.
Establishing cosmological constraints from cluster abundances requires two main steps: detecting clusters in large sky surveys, and assessing the masses (and redshifts) of cluster candidates \citep[see, \eg,][for a recent description of the methodology]{2023arXiv231012213B}.

Cluster detection can be performed through several means; \eg, identifying overdensities in the distribution of galaxies at optical and infrared wavelengths \citep[\eg,][]{2016ApJS..224....1R}; or observing a signal from the hot gas populating the intracluster medium (ICM), either from its \textit{bremsstrahlung} emission in the X-ray domain \citep[\eg][]{2024A&A...685A.106B}, or at millimeter wavelengths \citep[\eg,][]{2021ApJS..253....3H, 2024OJAp....7E..13B}, via its imprint on the cosmic microwave background (CMB) through the thermal Sunyaev-Zel'dovich (tSZ) effect \citep[][for a recent review]{1972CoASP...4..173S, 2019SSRv..215...17M}.
Subsequently, the mass calibration of cluster candidates may also follow two steps.
The first step is the absolute mass calibration of candidates, consisting of the estimation of cluster masses from observable cluster physical properties.
This is usually based on reconstructing the gravitational potential of the cluster, either through the gravitational lensing of background galaxies \citep[see, \eg,][for a review]{2020A&ARv..28....7U}, or by measuring the thermodynamic properties of the ICM and assuming hydrostatic equilibrium \citep[see, \eg,][for a review]{2019SSRv..215...25P}.
In the majority of cases, absolute mass calibration is not possible for all candidates in a cluster sample.
A second step, referred to as relative mass calibration, must then be employed, assessing the statistical relationship between cluster masses (obtained via absolute mass calibration) and survey observables, such as detection significance \citep[\eg][]{2024arXiv240102075B}, the integrated tSZ \citep[\eg,][]{2016A&A...594A..27P} or X-ray \citep[\eg,][]{2024arXiv240208458G} signal, or number of member galaxies \citep[\eg][]{2019MNRAS.482.1352M}.

As cosmological surveys probe the sky with increasing sensitivity, cluster samples become larger, and cluster cosmology becomes increasingly more precise, with the control of systematic uncertainties becoming a crucial challenge.
To this end, cluster cosmology has heavily relied on the use of cosmological simulations, which offer ideal synthetic datasets with a known underlying truth.
Simulations have been used for a wide range of applications: to calibrate the dependence of the halo mass function on cosmological parameters \citep[\eg,][]{2019ApJ...872...53M, 2020ApJ...901....5B}; to measure the correlations between different halo properties \citep[\eg,][]{2012MNRAS.426.2046A, 2021MNRAS.500.1029L} and their evolution in mass and redshift \citep[\eg,][]{2012ApJ...758...74B, 2023ApJ...944..221S}; to quantify the accuracy and precision of different means to estimate cluster properties from observational data \citep[such as cluster masses; \eg,][]{2011ApJ...740...25B, 2021MNRAS.502.5115G, 2021MNRAS.507.5671G, 2022MNRAS.515.3383D}; or to assess the performance of cluster detection algorithms from synthetic sky maps or galaxy catalogs \citep[\eg,][]{2019A&A...627A..23E, 2023MNRAS.522.4766Z, 2024OJAp....7E..13B}.

One of the main considerations in using simulations is their ability to make relevant predictions.
In particular, to be used to calibrate cosmological analyses, simulations must be able to predict observable quantities, which can be used as surrogates for observational data to calibrate analysis pipelines.
In the context of ICM-based studies, this leads to using hydrodynamic simulations, which include a wide variety of physical processes and naturally simulate the intracluster gas.
Remarkable advances have been made in the past decades \citep[see, \eg,][for reviews of the field]{2020NatRP...2...42V, 2023ARA&A..61..473C}.
Nonetheless, some challenges remain.

First, baryonic physics at small scales---such as feedback by active galactic nuclei, radiative cooling, or star formation---can have an impact on the properties of the intracluster gas, which needs to be taken into account.
As these processes take place at scales smaller than the typical resolution of simulations, their inclusion relies on empirical models.
Because these models impact many different scales, ensuring that they produce accurate realizations of the physical properties of interest---\eg the properties of the intracluster gas---without degrading others---\eg the stellar mass function---requires tremendous calibration efforts, and adds a layer of uncertainty to the resulting synthetic products.
Moreover, the intricacy of baryonic physics implies a high numerical complexity for the models used in the simulations, resulting in a significant computational cost to run large hydrodynamic simulations.
As a result, such simulations often have to compromise between delivering the large volumes and high resolutions needed for cosmological studies---such as, \eg, Magneticum\footnote{\url{http://www.magneticum.org}}, cosmo-OWLS \citep{2014MNRAS.441.1270L}, BAHAMAS \citep{2017MNRAS.465.2936M}, MilleniumTNG \citep{2023MNRAS.524.2539P}---and running many smaller volume simulations spanning a wide range of physical models---such as, \eg, the IllustrisTNG \citep{2018MNRAS.473.4077P} and CAMELS \citep{2021ApJ...915...71V} suites---with the recent FLAMINGO suite managing to achieve large volumes at high resolution for twelve different sets of sub-resolution models \citep{2023MNRAS.526.4978S}.

To circumvent these challenges, a common alternative is the use of gravity-only simulations with observables created in post-processing \citep[see, \eg,][for a recent review]{2022LRCA....8....1A}.
Gravity-only (GO) simulations evolve collisionless particles interacting only through gravity, greatly simplifying computations in comparison to hydrodynamic simulations.
As a result, these simulations are computationally cheaper, and can be used to create large volumes at high resolution \citep[\eg][]{2017ComAC...4....2P, 2019ApJS..245...16H, 2021ApJS..252...19H, 2021MNRAS.506.4210I, 2022ApJS..259...15F}, or designing suites with varying cosmological parameters \citep[\eg][]{2016ApJ...820..108H, 2019ApJ...875...69D, 2024arXiv240607276H}.
Their main drawback is that by only considering gravitational interactions, these simulations effectively treat all matter as being dynamically collisionless (they are often referred to informally as ``dark matter-only simulations''), and are therefore not able to directly produce baryonic observables.
To use their products to calibrate cosmological analyses, one must then use post-processing techniques, aimed at predicting the properties that baryons would have if they were in fact present in the simulations, as a function of the measurable dark matter properties.

Many mapping techniques for inferring baryonic distributions and properties from GO runs have been developed, often grouped under the umbrella terms of ``baryonification'' or ``baryon painting'', using different models and seeking to produce different observables.
In particular, the emulation of intracluster gas properties has driven the development of several models, including the early work of \citet{2005ApJ...634..964O, 2007ApJ...663..139B} and its extensions \citep[sometimes jointly referred to as ``baryon pasting'', \eg,][]{2010ApJ...725.1452S, 2017ApJ...837..124F, 2023MNRAS.519.2069O, 2023OJAp....6E..43K, 2024arXiv241100108L}; the ``baryonification'' algorithm of \citet{2015JCAP...12..049S} and its extensions \citep[\eg,][]{2019JCAP...03..020S, 2024arXiv240601672A, 2024arXiv240903822A}; halo models \citep[\eg,][]{2020A&A...641A.130M, 2024arXiv240118072P}; and deep learning approaches \citep[\eg,][]{2019MNRAS.487L..24T, 2023MNRAS.526.2812C}.
Combined with the ``painting'' of different foregrounds to the CMB, these models have been used to create high-quality maps of the millimeter-wave sky, such as those presented in \citet{2010ApJ...709..920S}, Websky \citep{2020JCAP...10..012S}, AGORA \citep{2024MNRAS.530.5030O}, or the HalfDome simulations \citep{2024arXiv240717462B}.
These datasets are very widely used to calibrate cosmological analyses based on CMB surveys, and are a cornerstone of millimeter-wave cosmology.

In this work, we introduce \picasso, a new baryon painting model focused on the thermodynamic properties of the intracluster gas.
The model combines a parameterized analytical mapping between the gravity-only matter distribution and gas properties with a machine learning approach to predict the parameters of said mapping from halo properties.
This combination allows \picasso to combine the advantages of analytical models---\ie, the physically-motivated approach and interpretability---with the numerical efficiency and flexibility of machine learning.
In addition, we designed \picasso to be particularly flexible regarding the inputs needed to make predictions of gas properties (both in terms of input halo properties and potential distribution used to make predictions), seeking to maximize its usability by the scientific community.
The model is trained on pairs of gravity-only and hydrodynamic simulations, ensuring realistic predictions of gas properties for individual halos, and allowing for the augmentation of gravity-only simulations to mimic hydrodynamic data products.
In addition to describing the model, we release it as a Python package, taking full advantage of \jax \citep{jax2018github} for differentiability and hardware acceleration, and include trained models, offering the capacity to generate high-quality synthetic datasets with no need for training.

This article is structured as follows.
In \S\ref{sec:model}, we describe the \picasso gas model, detailing how it can be used to predict intracluster gas thermodynamics from gravity-only halo properties.
We present the simulation suite used to train the model in \S\ref{sec:data}.
In \S\ref{sec:baseline}, we describe our ``baseline'' model, the first training of the \picasso model, optimized to reproduce non-radiative hydrodynamic gas properties from gravity-only halos with maximal information (\ie from an extensive list of halo properties, including mass assembly history).
The performance of the baseline model is presented in \S\ref{sec:baseline:results}.
In \S\ref{sec:beyond_baseline}, we re-train the \picasso model in more complex scenarios, \ie to predict gas thermodynamics from less information per halo, and in full-physics hydrodynamic simulations including sub-resolution models.
The numerical implementation of the \picasso model as a Python package is described in \S\ref{sec:code}.
We conclude and provide avenues for improvement in \S\ref{sec:end}.

\paragraph{Notations}
Quantities indexed with a $\Delta c$ ($\Delta m$) subscript, where $\Delta \in [200, 500]$, denote halo properties integrated within radius $R_{\Delta c}$ ($R_{\Delta m}$), corresponding to a halo-centric radius enclosing an average density $\Delta$ times greater than the critical density (mean matter density) of the Universe at the considered redshift.
$h$ is the reduced Hubble constant, defined as $h = H_0 / 100 \, {\rm km \cdot s^{-1} \cdot Mpc^{-1}}$.

\begin{figure*}
    \centering
    \includegraphics[width=0.99\linewidth]{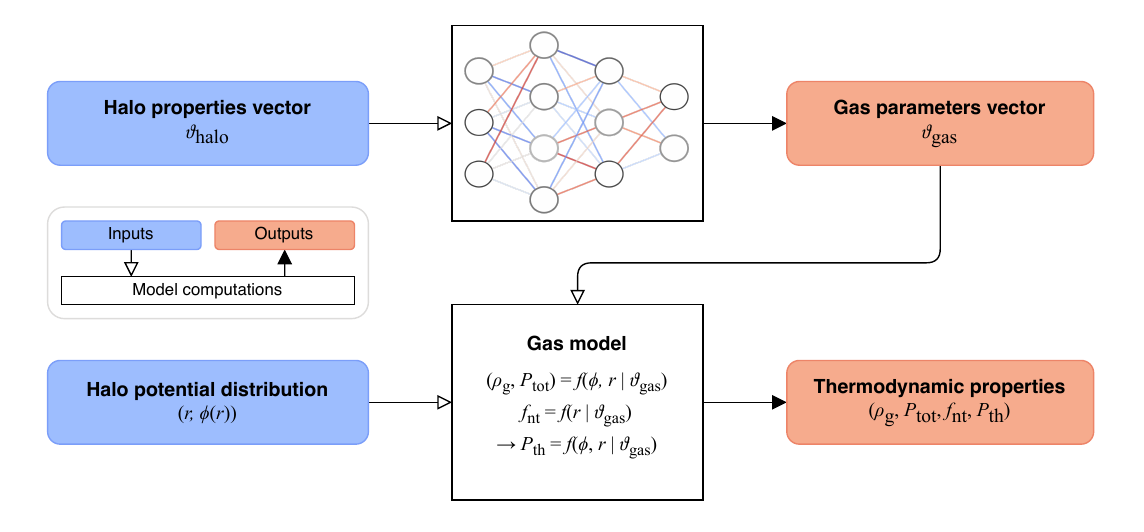}
    \caption{
        Schematic illustration of the \picasso gas model.
        For a given halo, the model inputs (blue) are its properties $\thetahalo$ and its gravitational potential distribution $\phi$.
        The model predictions (coral) are the gas model parameter vector $\thetagas$ and the resulting thermodynamic properties corresponding to the associated potential values.
    }
    \label{fig:model_schema}
\end{figure*}

\newpage
\section{The picasso gas model} \label{sec:model}

In this section, we describe the \picasso model for predicting ICM thermodynamics from the properties of halos in gravity-only simulations.
The overall workflow of the model is illustrated in \cref{fig:model_schema}.
As mentioned above, the model consists of two distinct parts:
\begin{itemize}[leftmargin=*]
    \item An analytical model mapping gas properties onto a gravitational potential distribution given a set of model parameters, $\thetagas$ (\S\ref{sec:model:gas_from_phi});
    \item A machine learning model predicting, for a given halo, the gas model parameter vector $\thetagas$ from a vector of halo properties as measured in a gravity-only simulation, $\thetahalo$ (\S\ref{sec:model:params_from_haloprops}).
\end{itemize}
We present these two components in the following subsections.

\subsection{Mapping gas properties on dark matter halos} \label{sec:model:gas_from_phi}

\subsubsection{Polytropic gas model}
Following previous work \citep[\eg,][hereafter \bpbc]{2010ApJ...725.1452S, 2023MNRAS.519.2069O, 2023OJAp....6E..43K}, we choose to model intracluster gas as obeying a polytropic equation of state and tracking the gravitational potential.
Specifically, we propose a modification of the polytropic model in an arbitrary potential of \citet{2005ApJ...634..964O}, in which the gas density $\rhog$ and total (thermal + non-thermal) pressure $\Ptot$ are written as:
\begin{align}
    \nonumber \frac{\rhog(\phi, \, r)}{500\rhocrit} &= \rho_0 \, \theta(\phi)^\frac{1}{\Gamma - 1}; \\
              \frac{\Ptot(\phi, \, r)}{P_{500c}}    &= P_0    \, \theta(\phi)^\frac{\Gamma}{\Gamma - 1},
    \label{eq:model_rho_P}
\end{align}
where $\rho_0$ and $P_0$ are the central gas density and total pressure respectively, $\Gamma$ is the gas polytropic index, and
\begin{equation}
    \theta(\phi) = 1 - \theta_0 \times \phi,
    \label{eq:model_theta}
\end{equation}
where $\phi$ is the normalized gravitational potential of the halo,\footnote{We define the normalized potential $\phi$ as the difference between the local potential and its value for the most bound halo particle, such that $\phi = 0$ at the bottom of the potential well and $\phi > 0$ everywhere else. This definition of $\phi$ corresponds to $\phi - \phi_0$ in the notation of \citet{2005ApJ...634..964O}.} and $\theta_0$ a parameter of the model, which we hereafter refer to as the polytropic normalization.

This model is equivalent to that of \citet{2005ApJ...634..964O} in the limit where
\begin{equation*}
    \theta_0 \rightarrow \frac{\Gamma - 1}{\Gamma} \frac{\rho_0}{P_0}.
\end{equation*}
Fixing the polytropic normalization to this specific value assumes that the gas is in hydrostatic equilibrium in the halo potential, \ie that the total gas pressure compensates for the gravitational collapse.
Thus, by allowing it to vary, we allow this equilibrium to be broken, at the expense of one additional model parameter. \\
Finally, in \cref{eq:model_rho_P}, the density and pressure are respectively normalized to the critical density at the redshift of the halo $\rhocrit(z)$, and to the characteristic pressure expected for a halo of mass $M_{500c}$ at redshift $z$ in the self-similar structure collapse scenario $P_{500c}$ \citep{2007ApJ...668....1N, 2010A&A...517A..92A}:
\begin{align}
    \label{eq:p500c}
    \frac{P_{500c}(M_{500c}, \, z)}{1.65 \times 10^{-3}} = E^{8/3}(z) \left[\frac{M_{500c}}{3 \times 10^{14} \, h_{70}^{-1} M_\odot}\right]^{2/3} & \\
    \nonumber {h_{70}^2 \, {\rm keV \cdot cm^{-3}}}, &
\end{align}
with $h_{70} = h/0.7$, and $E(z) = H(z) / H_0$.

\begin{figure}
    \centering
    \includegraphics[width=0.99\linewidth]{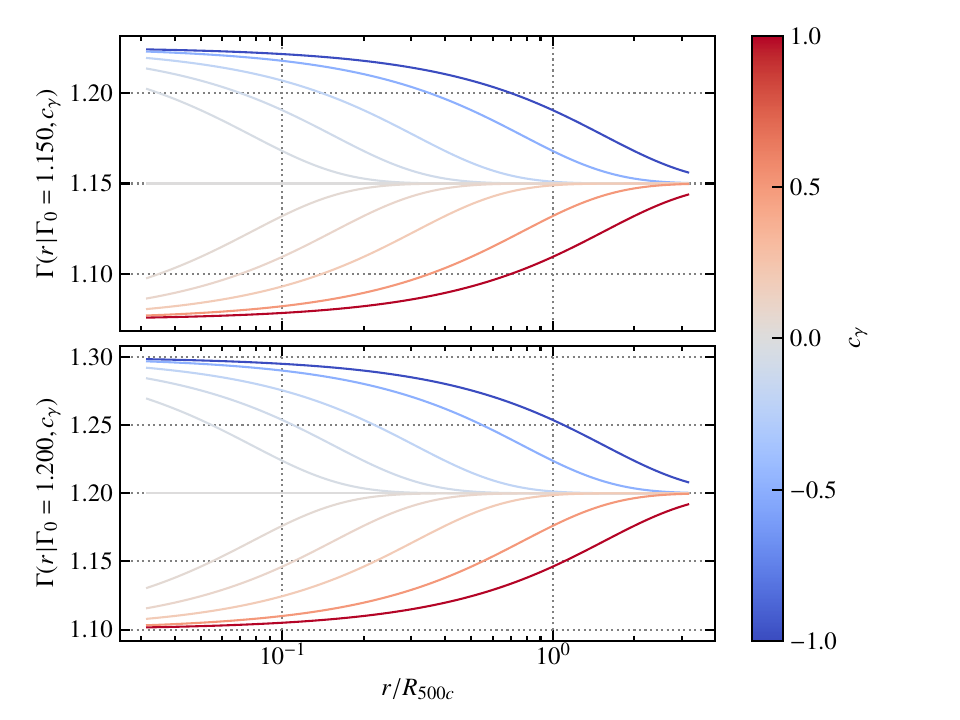}
    \caption{
        Radial evolution of the polytropic index $\Gamma(r)$ in the parameterization presented in \cref{eq:Gamma_r} for different values of $c_\gamma$ (colored lines).
        We show the evolution for $\Gamma_0 = 1.15$ (top) and $\Gamma_0 = 1.2$ (bottom).
    }
    \label{fig:gamma_r}
\end{figure}

In order to accommodate the different physical processes occurring at various cluster scales, we model the gas polytropic index $\Gamma$ as a function of the halo-centric radius $r$ via:
\begin{equation}
    \Gamma(r) = 
    \begin{cases}
    \begin{aligned}
        & \; 1 + (\Gamma_0 - 1) \frac{1}{1 + e^{-x}} & c_\Gamma > 0; \\
        & \; \Gamma_0 & c_\Gamma = 0; \\
        & \; \Gamma_0 + (\Gamma_0 - 1) \left(1 - \frac{1}{1 + e^{x}}\right) & c_\Gamma < 0, \\
    \end{aligned}
    \end{cases}
    \label{eq:Gamma_r}
\end{equation}
with
\begin{equation*}
    x = \frac{r}{c_\gamma \times R_{500c}}.
\end{equation*}
Here, $\Gamma_0$ is the asymptotic value for the adiabatic index as $x \rightarrow \infty$ and $c_\gamma$ is a shape parameter.
Both $\Gamma_0$ and $c_\gamma$ are free parameters of the model; the radial evolution of $\Gamma$ with this parameterization is illustrated in \cref{fig:gamma_r}.
Note that in most of this work, we fix $c_\gamma = 0$, corresponding to a constant polytropic index, $\Gamma(r) = \Gamma_0 \;\forall\, r$; an investigation of the impact of releasing this constraint is presented in \S\ref{sec:beyond:Gamma_r}.

\subsubsection{Non-thermal pressure fraction}
We are primarily interested in modeling the tSZ signal in clusters, which is sourced by the thermal pressure of the ICM, $\Pth \propto \rhog T$.
To derive this property from the total pressure in \cref{eq:model_rho_P}, we must also model the fraction of the total gas pressure that is due to non-thermal processes (in particular bulk motions within the ICM).
Several models have been proposed in the literature, from power laws of radius \citep[\eg,][]{2009ApJ...705.1129L, 2010ApJ...725.1452S, 2012ApJ...758...74B, 2012arXiv1204.1762B} to more complex formulations \citep[\eg,][]{2014MNRAS.442..521S, 2014ApJ...792...25N}.
Here, we propose to write the non-thermal pressure fraction as the sum of a constant plateau in the halo center and a power-law evolution with radius: 
\begin{equation}
    \fnt(r) = \Ant + (\Bnt - \Ant) \left(\frac{r}{\rfnt}\right)^{\Cnt},
    \label{eq:model_fnt}
\end{equation}
where $\Ant$ is the central non-thermal fraction plateau, $\Bnt$ is the non-thermal pressure fraction at $r=\rfnt$, and $\Cnt$ is the power-law dependence of the profile that dominates at large radii (\ie, $\fnt(r \ll \rfnt) \rightarrow \Ant$, and $\fnt(r \gg \rfnt) \sim r^{\Cnt}$).
The thermal gas pressure can then be obtained by combining \cref{eq:model_rho_P} and \cref{eq:model_fnt}:
\begin{equation}
    \Pth(r, \, \phi) = [1 - \fnt(r)] \times \Ptot(r, \, \phi).
    \label{eq:model_pth}
\end{equation}

\begin{figure*}
    \centering
    \includegraphics[width=0.99\linewidth]{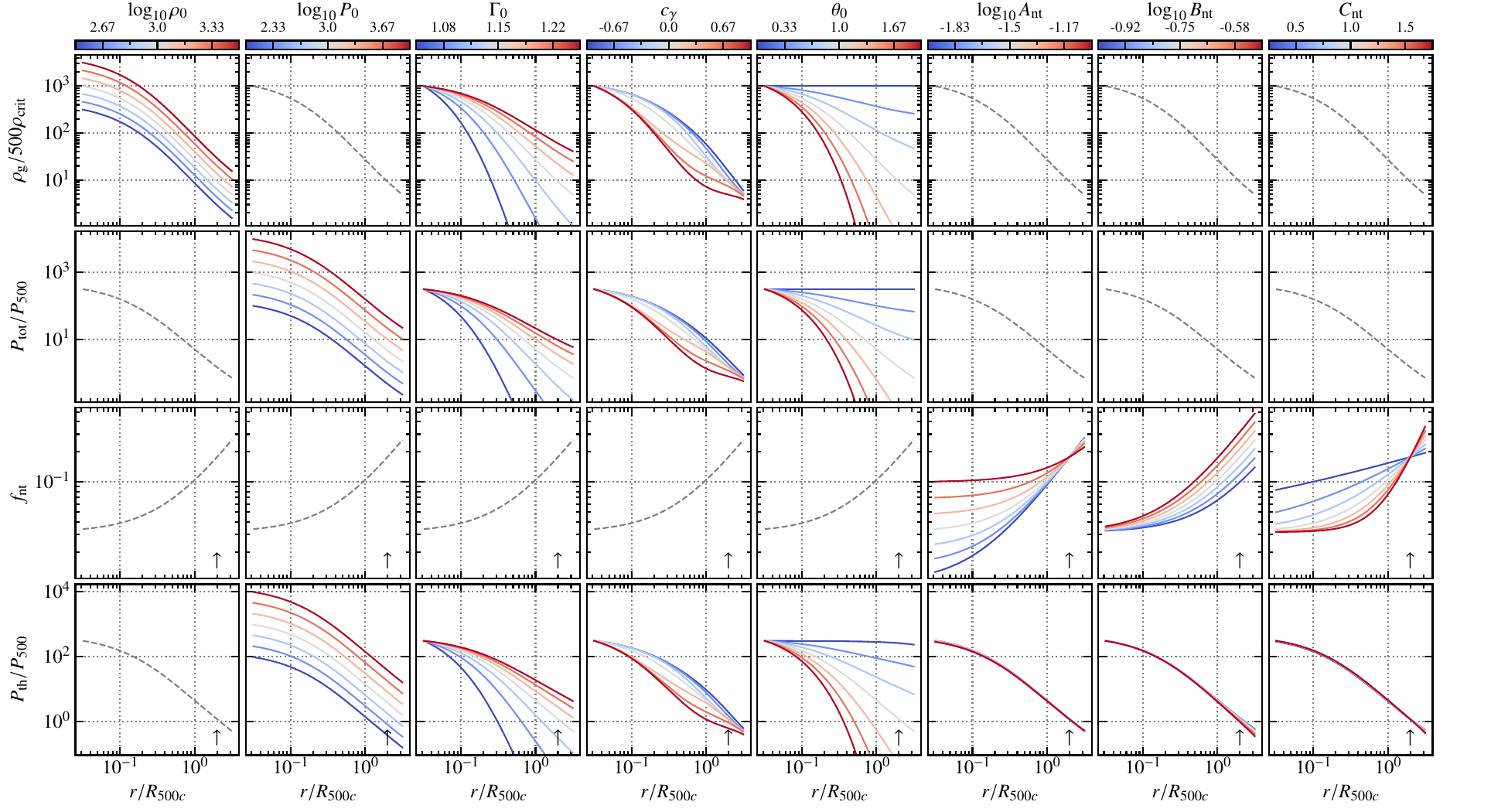}
    \caption{
        Sensitivity of the different thermodynamic properties in our model to the gas model parameters.
        From top to bottom, we show gas density, total pressure, non-thermal pressure fraction, and thermal pressure.
        From left to right, we show the impact of parameters $\rho_0$, $P_0$, $\Gamma_0$, $c_\gamma$, $\theta_0$, $\Ant$, $\Bnt$, and $\Cnt$.
        For each parameter, values range from low (blue) to high (red) according to the color code indicated at the top of the corresponding column.
        When varying a parameter, all others are fixed to the value corresponding to the central value of the interval over which they are varied.
        In each column, properties that are independent of the corresponding parameter are drawn as gray dashed lines corresponding to the central parameter value.
        Note that the thermal pressure (bottom row) is affected by the non-thermal pressure fraction (rightmost three columns), but the corresponding variation is of the order of a few percent and not visibly noticeable.
        For the bottom two rows, the vertical arrow shows the position of $\rfnt$, used to model the non-thermal pressure fraction.
    }
    \label{fig:impact_model_params}
\end{figure*}

\paragraph{Summary: the $\thetagas$ parameter vector}
Combined, \refmodeleqs provide a model that fully specifies the gas thermodynamic properties ($\rhog$, $\Ptot$, $\fnt$, $\Pth$) for a given value of gravitational potential $\phi$ and at a given distance $r$ of a halo center.
This model has eight free parameters, summarized in \cref{tab:params_thetagas}, which form the $\thetagas$ parameter vector.
The sensitivity of the different gas properties to the eight parameters is illustrated in \cref{fig:impact_model_params}.
We see that no single thermodynamic property is sensitive to all eight parameters; at most, parameters $\Gamma_0$, $c_\gamma$ and $\theta_0$ impact three of the four relevant properties.
We also note that the thermal pressure is sensitive to the most parameters (seven), although the impact of the non-thermal pressure fraction is much weaker than that of the other parameters.

\begin{table*}[t]
    \centering
    \begin{tabular}{c c c}
        \toprule
        Symbol & Meaning & Range\\
        \midrule
        \midrule
        $\log_{10} \rho_0$   & (log-scaled) Central normalized gas density & $(1.5, \, 5)$ \\
        $\log_{10} P_0$      & (log-scaled) Central normalized gas total pressure & $(0, \, 4.5)$ \\
        $\Gamma_0$                                  & Gas polytropic index limit as $r \rightarrow \infty$ & $(1, \, 1.4)$ \\
        $c_\gamma$                                  & Gas polytropic index shape parameter & $[0]$\footnote{\centering\footnotesize \;$(-1,\, 1)$ for the NR+$\Gamma(r)$ and SG+$\Gamma(r)$ models; see \S\ref{sec:beyond:Gamma_r}.} \\
        $\theta_0 / (10^{-6} \, {\rm km^2 s^{-2}})$ & Polytropic normalization & $(0, \, 2)$ \\
        $\log_{10} A_{\rm nt}$                      & (log-scaled) Central plateau of non-thermal pressure fraction & $(-4, \, 0)$ \\
        $\log_{10} B_{\rm nt}$                      & (log-scaled) Non-thermal pressure fraction at $r=\rfnt$ & $(-1.5, \, 0)$ \\
        $C_{\rm nt}$                                & Non-thermal pressure fraction profile power law index & $(0, \, 4)$ \\
        \bottomrule     
    \end{tabular}
    \caption{\normalfont
        Description of the components of the $\thetagas$ parameter vector.
        These parameters are used in \refmodeleqs to compute gas thermodynamic properties (see \S\ref{sec:model:gas_from_phi}).
        The last column indicates the range the parameters are allowed to vary within for the baseline model---see \S\ref{sec:baseline:nn}.
    }
    \label{tab:params_thetagas}
\end{table*}

\subsection{Predicting gas model parameters} \label{sec:model:params_from_haloprops}

To use the model presented above to predict gas properties, we must compute an estimate of the parameter vector $\thetagas$.
Previous studies based on similar models have used the prescription of \citet{2005ApJ...634..964O}, modeling a physical transformation of the gas to solve for some of the parameters, and fixing the others \citep[\eg,][]{2010ApJ...725.1452S, 2023MNRAS.519.2069O} or adjusting them at the population level, using observations \citep[\eg,][]{2017ApJ...837..124F} or hydrodynamic simulations (see, \eg, \bpbc).
These approaches, while yielding good results and leading to very broadly used synthetic datasets \citep[\eg,][]{2010ApJ...709..920S}, can prove relatively costly from a computational standpoint, in particular when trying to apply them to arbitrary potential shapes from the dark matter particles of a gravity-only simulation \citep[\eg,][]{2023MNRAS.519.2069O}.
Moreover, they assume a scenario in which the intracluster gas undergoes a physical transformation from following the dark matter exactly to a polytropic equation of state while conserving energy and boundary conditions.
While this assumption makes the model very attractive for its physicality, it relies on a simplified model of halo growth, and may not make optimal use of the wealth of information contained in cosmological simulations.

We propose a different approach, in which the model parameter vector $\thetagas$ is determined using machine learning.
Specifically, we design a neural network to predict the full $\thetagas$ vector corresponding to a given halo given a vector of its properties.
This vector---hereafter denoted $\thetahalo$---can contain a multitude of halo properties measurable in gravity-only simulations, such as mass, concentration, disturbance indicators, or of summary statistics of the mass accretion history, which are known to contain valuable information on halos and to correlate tightly to their observed properties \citep[see, \eg,][]{2014MNRAS.442..521S, 2021MNRAS.500.1029L}.
We will discuss examples of input data vectors---and investigate their respective predictive power---in \S\ref{sec:baseline:thetahalo} and \S\ref{sec:beyond_baseline}, and discuss the implementation of a network and its training in \S\ref{sec:baseline:nn} and \S\ref{sec:baseline:loss}, respectively.

\section{Training data} \label{sec:data}

Our model optimization strategy follows that presented in \bpbc, in which, for a given gravity-only halo, the expected hydrodynamic properties are those of its counterpart in a hydrodynamic simulation with the same initial conditions.
This section describes the simulations used to train our models.

\subsection{Simulations} \label{sec:data:576}

The simulation dataset was generated using the Hardware/Hybrid Accelerated Cosmology Code (\HACC).
This framework includes sophisticated gravity-only \citep{habib_hacc_2016} and hydrodynamics solvers \citep{frontiere_simulating_2023}, optimized for high performance on modern supercomputing platforms, including GPU hardware acceleration.
Three simulations were performed using identical initial conditions with increasingly detailed physics modeling: a gravity-only simulation, a non-radiative hydrodynamics simulation, and a ``subgrid'' simulation including astrophysical feedback and galaxy formation models.
The simulations trace 2304$^3$ dark matter particles in a volume of $V=(576\,h^{-1}{\rm Mpc})^3$, with the hydrodynamics suite further evolving an equal number of baryon particles subject to gas physics\footnote{The corresponding total matter particle mass resolution is $1.34 \times 10^9 \, h^{-1}M_\odot$ for the gravity-only simulation, and dark matter and baryon mass resolutions are $1.13\times 10^9 \, h^{-1}M_\odot$ and $2.12\times 10^8 \, h^{-1}M_\odot$, respectively for the gas simulations.}.
The force resolution of all three simulations is $10\,h^{-1} {\rm kpc}$. 
The subgrid simulation includes models for radiative cooling and ultraviolet background heating, star formation, supernova feedback and galactic winds, chemical enrichment, and active galactic nuclei feedback. The individual model parameters were calibrated to observations, such as the galaxy stellar mass function and the black hole stellar mass relation.
More detail on the simulations and parameterizations can be found in \citet[\S2]{2024arXiv240700268S}.

\HACC utilizes extensive GPU-accelerated in situ and post-processing analysis pipelines, which generate detailed structure formation data products.
These simulation outputs include comprehensive halo and galaxy catalogs, in addition to full merger tree histories, as well as substructure tracking utilizing halo cores \citep{2021ApJ...913..109S, 2023OJAp....6E..24K, 2024arXiv240700268S}.
Halos are identified using a Friends-of-Friends (FOF) halo finder on dark matter particles with a specified linking length of $b=0.168$ times the inter-particle separation, and a center defined to be the gravitational potential minimum.
Spherical overdensity halos are then constructed from the center, including all particle species when applicable.
For details of the specific outputs see \citet{rangel2017,2021ApJS..252...19H, 2024arXiv240700268S}.
A summary of the individual properties utilized for the training of the \picasso model is listed in \S~\ref{sec:baseline:thetahalo}.

\subsection{Halo matching} \label{sec:data:halo_match}

For the training of our models, we restrict ourselves to the last snapshot of the simulation ($z=0$), and to halos with masses $M_{500c} > 10^{13.5} \, h^{-1}M_\odot$, roughly corresponding to a mass scale ranging from massive groups to clusters.
For each halo satisfying this mass cut in the gravity-only volume, we search for a counterpart in the hydrodynamic volumes.
Specifically, a halo in a hydrodynamic run is accepted as a counterpart to a gravity-only halo if it meets the two following requirements:
\begin{itemize}[leftmargin=*]
    \item Its friends-of-friends center is located within the gravity-only halo radius $R_{500c}$;
    \item Its mass $M_{500c}$ is within $20\%$ of that of the gravity-only halo (allowing us to avoid matching halos with subhalos).
\end{itemize}
With these two criteria, we find a suitable match for $\sim 95\%$ of gravity-only halos in both the non-radiative and subgrid runs, respectively resulting in sample sizes of 8,220 and 8,306 halos.

\subsection{Radial profile estimation} \label{sec:data:profiles}

Per \refmodeleqs, the \picasso gas model predicts gas properties based on the local gravitational potential and distance from the cluster center.
This means that they can be computed in a variety of different ways, such as projections on three-dimensional grids, or directly at the particle positions from the output of the $N-$body simulation.
In order to limit memory requirements during the training stage, we use one-dimensional radial profiles, allowing us to make predictions for a large number of halos at a time.
We choose a logarithmically-spaced radial binning, with bin edges:
\begin{align}
    \nonumber r/R_{500c} = [ & 0,\, 0.1,\, 0.134,\, 0.195,\, 0.271,\, 0.379,\, \\
                             & 0.528,\, 0.737,\, 1.028,\, 1.434,\, 2,\, 3 ].
    \label{eq:rbins}
\end{align}

For each halo, we first combine \cref{eq:rbins} and the halo radius $R_{500c}^{\rm GO}$ in the gravity-only run to compute the radial edges corresponding to the gravity-only run.
We then use these edges to define concentric spherical shells around the gravity-only FOF halo center, and compute the radial potential profile, $\phi_{\rm GO}(r)$, as the average value of the normalized potential at the locations of all particles within each shell.

The thermodynamic profiles are evaluated similarly, starting with the non-radiative simulation.
First, \cref{eq:rbins} is used to compute the shell edges for the non-radiative run, using the corresponding halo radius\footnote{Note that because halos may have slightly different masses in each simulation flavor, their radii $R_{500c}$ also vary slightly. Therefore, for each flavor, we compute the bin edges using the halo radii specific to that simulation flavor.} $R_{500c}^\NR$.
We then measure the radial profiles of the following thermodynamic properties around the center of the halo in the non-radiative run by computing, for each shell:

\begin{itemize}[leftmargin=*]
    \item The normalized gas density:
    \begin{equation}
        \rhog^\NR(r) = \frac{1}{\vshell(r)} \sum_{i=1}^{N} m_{\rm g,i},
    \end{equation}
    where $\vshell$ is the volume of the corresponding shell, and the sum runs over all $N$ gas particles $i$ of mass $m_{\rm g,i}$ within the radial shell (note that the gas particles have constant mass in the non-radiative run, but can vary in the subgrid run---see\S\ref{sec:data:576});
    \item The normalized thermal pressure:
    \begin{equation}
        \Pth^\NR(r) = \frac{2}{3} \rhog(r) \left< u \right>,
    \end{equation}
    where $\left< u \right>$ is the mass-averaged thermal energy of gas particles within the shell;
    \item The normalized total (thermal + kinetic) pressure:
    \begin{equation}
        \Ptot^\NR(r) = \Pth^\NR(r) + \frac{\rhog(r)}{3}  \left< \delta v \cdot \delta v \right>,
    \end{equation}
    where $\left< \delta v \cdot \delta v \right>$ is the average velocity fluctuation of gas particles within the shell, measured with respect to the gas center of mass and mass-averaged velocity within the halo radius;
    \item The fraction of non-thermal pressure:
    \begin{equation}
        \fnt^\NR(r) = 1 - \frac{\Pth(r)}{\Ptot(r)}.
    \end{equation}
\end{itemize}
To ensure that these profiles are comparable with the model predictions defined in \refmodeleqs, we normalize them similarly and compute:
\begin{equation}
    \tilde{\rho}_{\rm g}^\NR = \frac{\rhog^\NR}{500\rhocrit} \;;\;
    \tilde{P}_{\rm tot}^\NR  = \frac{\Ptot^\NR}{P_{500c}^\NR}    \;;\;
    \tilde{P}_{\rm th}^\NR   = \frac{ \Pth^\NR}{P_{500c}^\NR},
    \label{eq:tildes}
\end{equation}
where $P_{500c}^\NR = P_{500c}(M_{500c}^\NR, z)$, per \cref{eq:p500c}.

The same procedure is then used to estimate the thermodynamic profiles in the subgrid run to compute $\tilde{\rho}_{\rm g}^\SG$, $\tilde{P}_{\rm th}^\SG$, $\tilde{P}_{\rm tot}^\SG$, and $\fnt^\SG$ (see \S\ref{sec:beyond:fullhydro}).

\section{Training picasso: the baseline model} \label{sec:baseline}

Summarizing what has been described so far, \S\ref{sec:model} provided a description of the gas model, making predictions of gas thermodynamics from an input vector of halo properties and a spatial distribution of the gravitational potential, and \S\ref{sec:data} described the data products available for training, \ie, the properties of halos and of their potential distribution in a gravity-only simulation, and the expected gas properties for these halos in two hydrodynamic simulations (non-radiative and subgrid hydrodynamics).
In this section, we describe the first training of the \picasso gas model, hereafter referred to as our ``baseline'' model.

\subsection{Input vector properties $\thetahalo$} \label{sec:baseline:thetahalo}

\begin{table*}[t]
    \centering
    \begin{tabular}{c c c c}
        \toprule
        Symbol & Meaning & Compact? & Minimal? \\
        \midrule
        \midrule
        $\log_{10} (M_{200c} / 10^{14} \, h^{-1}M_\odot)$ & (log-scaled) Halo mass & \yay & \yay \\
        $c_{200c}$  & Halo concentration & \yay & \yay \\
        \midrule
        $\Delta x / R_{200c}$   & Normalized offset between center of mass and potential peak             & \yay & \nay \\
        $c_{\rm acc.}/c_{200c}$ & Ratio between accumulated mass and NFW fit concentrations               & \yay & \nay \\
        $c_{\rm peak}/c_{200c}$ & Ratio between differential mass profile peak and NFW fit concentrations & \yay & \nay \\
        \midrule
        $e$ & Halo ellipticity, \cref{eq:ell_pro} & \yay & \nay \\
        $p$ & Halo prolaticity, \cref{eq:ell_pro} & \yay & \nay \\
        \midrule
        $a_{\rm \rm lmm}$ & Scale factor of last major merger & \nay & \nay \\
        $a_{25}$  & Scale factor at which $M=0.25 \times M_{z=0}$ & \nay & \nay \\
        $a_{50}$  & Scale factor at which $M=0.50 \times M_{z=0}$ & \nay & \nay \\
        $a_{75}$  & Scale factor at which $M=0.75 \times M_{z=0}$ & \nay & \nay \\
        $\dot{M}$ & Mass accretion rate between last two redshift snapshots & \nay & \nay \\
        \bottomrule     
    \end{tabular}
    \caption{\normalfont
        Description of the components of the $\thetahalo$ parameter vector used in the baseline model presented in \S\ref{sec:baseline}.
        The last two columns denote whether the properties are included in the compact  (second rightmost) and minimal (rightmost) models, presented in \S\ref{sec:beyond:compact} and \S\ref{sec:beyond:minimal}, respectively.
    }
    \label{tab:params_thetahalo}
\end{table*}

The first choice needed to perform the training of the \picasso model lies in specifying the set of halo properties used in the input data vector.
For the baseline model, we design this vector to include maximal information on the halo, including most\footnote{Note that we avoid including highly correlated properties, such as different definitions of halo mass or halo mass proxies, to avoid network confusion due to colinear inputs.} properties from the halo catalog, as well as assembly history information derived from the halo merger trees.
The components of the input vector $\thetahalo$ are:

\begin{itemize}[leftmargin=*]
    \item Halo mass $M_{200c}$ and concentration $c_{200c}$, measured by fitting a Navarro-Frenk-White \citep[NFW,][]{1997ApJ...490..493N} model on the halo density profile.
    \item Disturbance: we use three indicators known to contain information on the relaxation state of halos.
        (a) the normalized offset between the center of mass and potential peak $\Delta x / R_{200c}$;
        (b) the ratio between concentration measured from the accumulated mass profile and an NFW fitting, $c_{\rm acc.}/c_{200c}$;
        (c) the ratio between concentration measured from the differential mass profile peak and NFW fitting, $c_{\rm peak}/c_{200c}$.
        More details on these indicators in the context of \HACC simulations can be found in \citet{2018ApJ...859...55C}, \S3.
    \item Halo shape: We use halo ellipticity, measuring a halo's deviation from sphericity, and prolaticity, quantifying the extent to which a halo is oblate (disk-shaped) or prolate (cigar-shaped).
    They are defined through the halo's semi-axes $a, b, c$, where $a \geqslant b \geqslant c > 0$, as:
    \begin{align}
        \nonumber e &= \frac{1}{2L} \left(1 - (c/a)^2 \right); \\
        p &= \frac{1}{2L} \left( 1 - 2(b/a)^2 + (c/a)^2 \right),
        \label{eq:ell_pro}
    \end{align}
    where $L  = 1 + (b/a)^2 + (c/a)^2$.
    For a given halo, $e$ ranges between 0 (spherical) and 1/2 (non-spherical), and $p$ between $-e$ (oblate) and $e$ (prolate).
    The semi-axes are computed from the eigenvalues of the reduced inertia tensor of the halo particles \citep[\eg][]{2006MNRAS.367.1781A, 2021MNRAS.500.1029L}.
    \item Mass assembly history: we use the scale factors $(a_{25}, \, a_{50}, \, a_{75})$ at which the halo has achieved $25\%$, $50\%$ and $75\%$ of its final mass, respectively; as well as the instanteous mass accretion rate of the halos, $\dot{M}$, between the last two redshift snapshots of the simulation, and the scale factor at the last major merger, $a_{\rm lmm}$, defined as the scale factor of the Universe when a given halo underwent its last merger with mass ratios greater than 0.3.
\end{itemize}
A summary of the notations and definitions for the components of $\thetahalo$ can be found in \cref{tab:params_thetahalo}.
Alternative models, relying on input vectors containing subsets of these properties, will be discussed in \S\ref{sec:beyond_baseline}.

\subsection{Neural network architecture} \label{sec:baseline:nn}

As mentioned in \S\ref{sec:model:params_from_haloprops}, we use a machine learning model to make predictions for the parameter vector $\thetagas$ from $\thetahalo$.
Specifically, for the baseline model, we choose a fully-connected neural network.
First, the input vector $x$ is defined as a linear rescaling of $\thetahalo$: for each feature $i$,
\begin{equation}
    x_i = \frac{\thetahaloi - {\rm min}(\thetahaloi)}{{\rm max}(\thetahaloi) - {\rm min}(\thetahaloi)},
\end{equation}
where ${\rm min}(\thetahaloi)$ and ${\rm max}(\thetahaloi)$ are the minimum and maximum values found in the dataset for the component $i$ of the $\thetahalo$ vector, ensuring that, for each component, the values of $x_i$ are contained between 0 and 1.
The input layer uses 12 features, corresponding to the components of the $x$ vector, and uses a scaled exponential linear unit (SELU) activation function.
It is followed by two fully-connected hidden layers, each with 32 features, and also using a SELU activation function\footnote{Different alternatives were tested, using deeper and wider architectures, as well as different activation functions; all resulting in similar results. Shallower and narrower networks were also investigated, yielding slightly worse accuracy, thus we chose the simplest architecture that achieved the best observed performance.}.
The output layer contains 8 features, corresponding to the components of the $\thetagas$ vector, and uses a sigmoid activation function.
This means that the network produces raw outputs $y$ that are bounded between 0 and 1; these outputs are then linearly rescaled:
\begin{equation}
    \thetagasi = y_i \times \left[ \thetagasi^{\rm max} - \thetagasi^{\rm min} \right] + \thetagasi^{\rm min},
\end{equation}
where the minimum and maximum values for each parameters are reported in \cref{tab:params_thetagas}.
This is equivalent to setting soft bounds on the parameters of the gas model, where the bounds are fixed \textit{a priori} to ensure a large parameter range while avoiding values resulting in numerical errors (\eg $\Gamma_0 = 1$).
We emphasize that, as noted in \S\ref{sec:model:gas_from_phi}, for the baseline model, we fix $c_\gamma = 0$, meaning the polytropic index of the gas is kept constant with radius; we explore models with $c_\gamma \in (-1, 1)$ in \S\ref{sec:beyond:Gamma_r}.

\subsection{Forward modeling of gas properties} \label{sec:baseline:fwmod}

The training of the model is based on comparing the gas properties predicted by \picasso for a gravity-only halo with those of counterparts in a matched hydrodynamic simulation.
For each halo, the input vector $\thetahalo$ is used to predict $\thetagas$ as described in \S\ref{sec:baseline:nn}.
The predicted parameter vector is then used in \refmodeleqs, along with the gravity-only potential profile $\phi_{\rm GO}(r)$, to compute the radial profiles of thermodynamic properties $Y$:
\begin{equation}
    Y_1 = \tilde{\rho}_{\rm g} \;;\quad
    Y_2 = \tilde{P}_{\rm tot} \;;\quad
    Y_3 = \fnt \;;\quad
    Y_4 = \tilde{P}_{\rm th},
    \label{eq:Y_props}
\end{equation}
with densities and pressures normalized similarly to the profiles extracted from the hydrodynamic simulation (see eq. \ref{eq:tildes}).

\subsection{Loss function and model training} \label{sec:baseline:loss}

Section \S\ref{sec:baseline:fwmod} described how we make predictions for the thermodynamic profiles of gravity-only halos using forward modeling.
The predicted profiles are then compared to the target data (\ie the profiles from the non-radiative hydrodynamic simulation) through a mean absolute error (MAE) loss function.
We selected the MAE loss after experimenting with various alternatives---including the more conventional mean squared error and median absolute error, as well as more sophisticated options like Huber and log-cosh loss functions---as it delivered the best balance between bias and variance in predicting thermodynamic profiles.

For a given set of weights and biases of the fully-connected neural network, hereafter denoted $\thetanet$, the loss value is written as:
\begin{equation}
    L(\thetanet) = \frac{1}{4} \sum_{i=1}^{4} \frac{1}{N_{\rm bins}}\sum_{j=1}^{N_{\rm bins}} \frac{1}{N_{\rm halos}} \sum_{k=1}^{N_{\rm halos}} \left| \frac{Y_{i, j, k}^{\rm pred} - Y_{i, j, k}^\NR}{Y_{i, j, k}^\NR} \right|,
    \label{eq:loss}
\end{equation}
where the triple summation runs over all $N_{\rm halos}$ halos $k$ in our dataset, the $N_{\rm bins}$ radial bins $j$ within a radial range \fullrrange, and over the four properties $Y_i$ of interest defined in \cref{eq:Y_props}, for which ``pred'' and ``NR'' superscripts denote \picasso predictions and non-radiative hydrodynamic profiles, respectively.
The choice of radial range is motivated in the inner region by the difficulty of modeling cluster cores---not only because of the importance of sub-resolution physics in these regions, but also because of the force softening used in the simulation---and by the low particle counts in the outskirts.
By limiting ourselves to $r > 0.1\times R_{500c}$, we ensure that we exclude regions within less than four times the force softening length from the halo center for all objects in our sample.
The outer limit, $r < 2\times R_{500c}$, is chosen by visually inspecting halo profiles and cutting out regions found to present systematically noisy profiles.

To train our model, we seek the set of network hyperparameters $\thetanet$ that minimizes \cref{eq:loss}---\ie, that maximize the agreement between \picasso predictions and hydrodynamic simulations for every halo in our sample.
To do so, we first divide our sample in three: a \textit{training} set, comprising $80\%$ of the dataset (6576 halos); a \textit{validation} set, and a \textit{testing} set, each including $10\%$ of the halos (822).
By writing the model predictions and the loss function using \jax \citep[][see \S\ref{sec:code}]{jax2018github}, we are able to automatically differentiate it with respect to the components of $\thetanet$.
This allows us to use efficient gradient-based optimization of the hyperparameters; specifically, we use the \texttt{adam} optimizer \citep{kingma2014adam} with an exponentially-decaying learning rate\footnote{Similar results were obtained with a constant learning rate, albeit resulting in a slower convergence.}, starting at $10^{-2}$ and plateauing at $10^{-4}$.
The $\thetanet$ hyperparameter vector is initialized at random.
The corresponding loss function is computed on the training set using \cref{eq:loss}, as well as its gradients with respect to each component of $\thetanet$.
The values of the hyperparameters $\thetanet$ are then changed in the direction of the gradients.
This process is repeated for a fixed (10,000) number of steps.
At each step of the minimization, the loss function is also computed for the validation set, and we store the values of hyperparameters $\thetanet$, of the training loss, and of the validation loss.
The testing set is not used during the training stage, and is saved to assess the performance of the model, as presented in \S\ref{sec:baseline:results}.

\begin{figure}[t]
    \centering
    \includegraphics[width=0.95\linewidth]{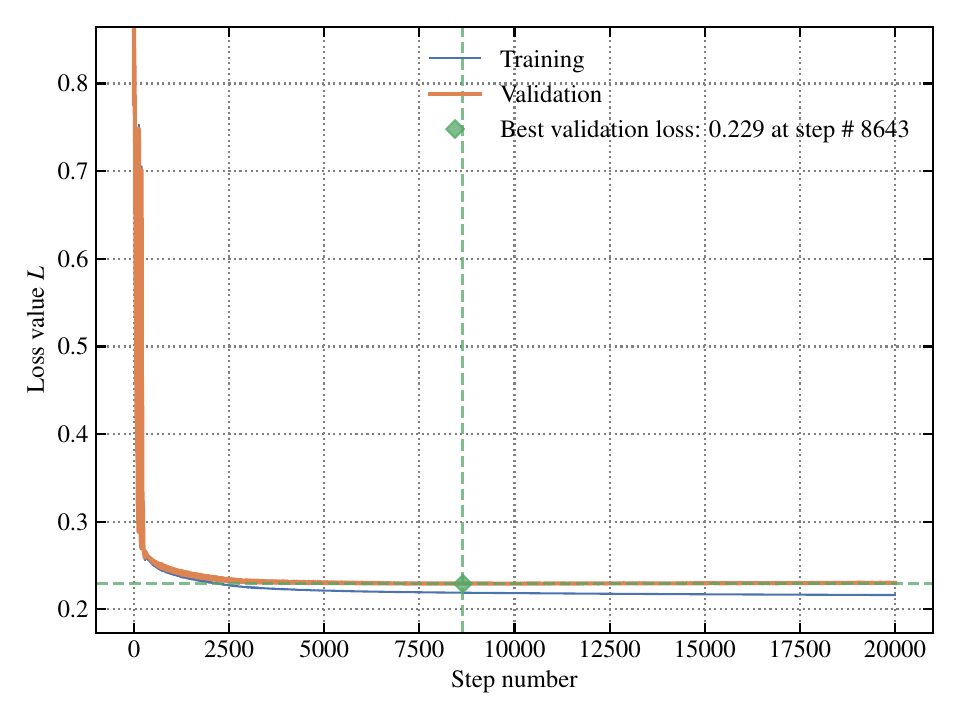}
    \caption{
        Loss function evolution during the baseline model training, for the training (blue) and validation (orange) sets.
        The green diamond and dashed lines show the position of the minimum validation loss, chosen as the optimal point in the training.
    }
    \label{fig:loss_baseline}
\end{figure}

The evolution of the loss values for the training and validation datasets is shown in \cref{fig:loss_baseline}.
We choose the set of optimized hyperparameters as the set that corresponds to the minimum of the validation loss curve, allowing to avoid overtraining the model to the training set.
This optimal state is reached in less than 9000 steps.
After this point, the training loss continues decreasing while the validation loss slightly increases back, showing the model entering the overtraining regime.
The set of optimized hyperparameters is saved, and is made available (see \S\ref{sec:code} for more information).

\section{Baseline model training results} \label{sec:baseline:results}

Once the model has been trained such that the optimal set of neural network hyperparameters $\thetanet$ that minimizes the loss in \cref{eq:loss} for the validation dataset has been found, we can assess its ability to make accurate and precise predictions of gas thermodynamic properties.
To that end, we use the testing dataset, the subset of 10\% of the halos which belong in neither of the training or validation sets, such that their properties have not played any role in the training phase.

\subsection{Accuracy and precision} \label{sec:baseline:reldiffs}

\begin{figure*}
    \centering
    \includegraphics[page=1, width=0.99\linewidth]{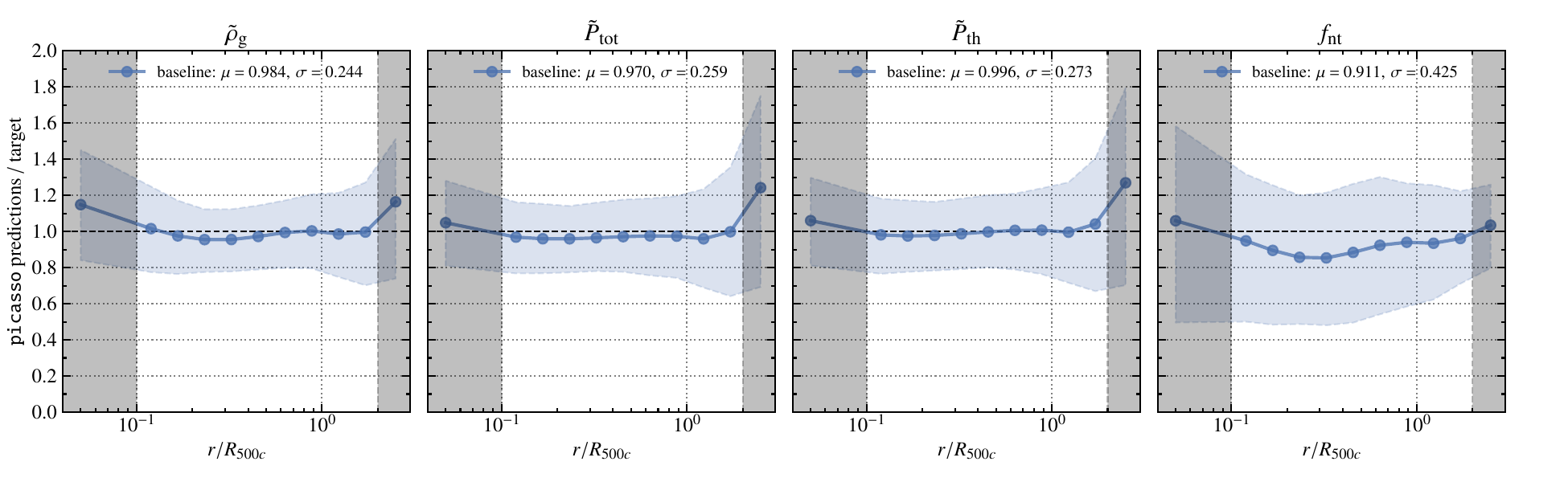}
    \caption{
        Distribution of the ratios between the \picasso predictions and the target data (from the non-radiative hydrodynamic simulation) for the radial profiles of the ICM thermodynamic properties for the baseline model:
        from left to right, gas density, total pressure, thermal pressure, and non-thermal pressure fraction.
        Results are shown for the testing set (\ie data entirely unused during the training process).
        The solid line shows the average ratio, while the shaded region represents the interval between the 16th and 84th percentile.
        The gray-shaded regions show the radial range not considered during the training.
        In each panel, the legend reports the mean $\mu$ and standard deviation $\sigma$ of the ratios within the radial range of interest \fullrrange.
        We see that \picasso can predict gas density and pressure with a few-percent average accuracy on the entire radial range.
    }
    \label{fig:perf:reldiffs_baseline}
\end{figure*}

\begin{figure*}
    \centering
    \includegraphics[width=0.99\linewidth]{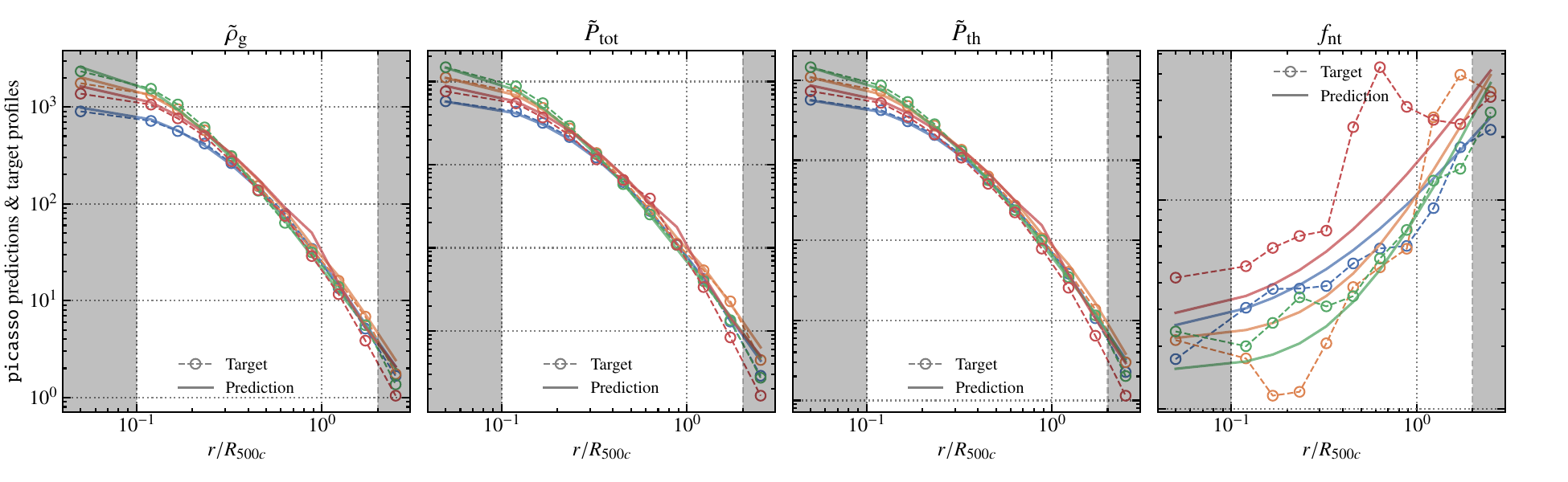}
    \caption{
        Comparison between \picasso predictions and target data (from the non-radiative hydrodynamic simulation) for the radial profiles of the ICM thermodynamic properties for the baseline model:
        from left to right, gas density, total pressure, thermal pressure, and non-thermal pressure fraction.
        We show results for four halos picked at random in the testing set, represented by the different colors.
        For each halo, the target data is shown as open circles connected by dashed lines, while the \picasso prediction is shown with a thick solid line.
        The gray-shaded regions show the radial range not considered during the training.
    }
    \label{fig:perf:profiles_example}
\end{figure*}

For each halo in the testing set, we use the trained model to predict the four properties of interest, following \S\ref{sec:baseline:fwmod}, and the ratio between these predictions and the target data (\ie thermodynamic profiles of halos in the non-radiative hydrodynamic simulation).
Results are shown in \cref{fig:perf:reldiffs_baseline}.
On the radial range of interest, \fullrrange, we can see that on average, the model predictions are biased at the percent level for the gas thermal pressure, and at the few-percent level for the gas density and total pressure.
The non-thermal pressure fraction is not recovered as accurately, with an average bias of 8\%, reaching up to 18\% at $r \simeq 0.3 R_{500c}$.

We also show the dispersion of the ratios between predictions and target data in \cref{fig:perf:reldiffs_baseline}.
We can see that the thermal pressure predictions have a scatter around the target data of about 20\% in the radial range $r/R_{500c} \in [0.1, \, 1]$, and increasing at larger radii.
We observe a similar behavior---with smaller scatter values---for the gas density and total pressure.
Again, the non-thermal pressure fraction is not recovered as precisely (with a scatter around 40\% for \fullrrange), although we note that the radial trend is inverted (the non-thermal pressure fraction is predicted more precisely in the outskirts).

Focusing on the thermal pressure, which is of particular interest as it sources the thermal Sunyaev-Zel'dovich signal of clusters, we see that the \picasso baseline model is very competitive, achieving accuracy and precision similar to algorithms like baryon pasting \citep[\eg][]{2023MNRAS.519.2069O} optimized to reproduce halo gas properties in non-radiative hydrodynamic simulations---see, \eg, \bpbc, reporting a $2\%$ bias and $\sim 20\%$ scatter in the range $r/R_{500c} \in \left[0.25, \, 1.25\right]$.
Moreover, using calibrated baryon pasting, \bpbc found that this level of precision resulted in a fractional increase of a few percent in the intrinsic scatter in the $Y_{500c} | M_{500c}$ scaling relation compared to non-radiative hydrodynamic simulations.
Given the comparative simplicity of the model (replacing the resolution of systems of equations on large data volumes by a fully-connected, pre-trained neural network) and the resulting computational speed-up (see discussion in \S\ref{sec:code:benchmark}), this is highly promising for the use of \picasso-generated synthetic data products in cluster cosmology, which will be assessed further in future works.

For illustration purposes, we also show the target data and predicted profiles for four halos randomly selected from the testing set in \cref{fig:perf:profiles_example}.
We see that the predictions for pressure and density agree with the profiles from the hydrodynamic simulation.
In contrast, the non-thermal pressure fraction predictions do not match the target data as closely, reflecting the higher scatter observed in \cref{fig:perf:reldiffs_baseline}.
This can be interpreted as a lack of sufficient information in halo summary statistics to make robust and precise predictions of non-thermal pressure fraction, which is expected, given the stochastic nature of kinetic pressure contributions (being due to, \eg, bulk motions, shocks, and turbulence).

\subsection{Parameter correlations}

\begin{figure*}
    \centering
    \includegraphics[page=1, width=0.99\linewidth]{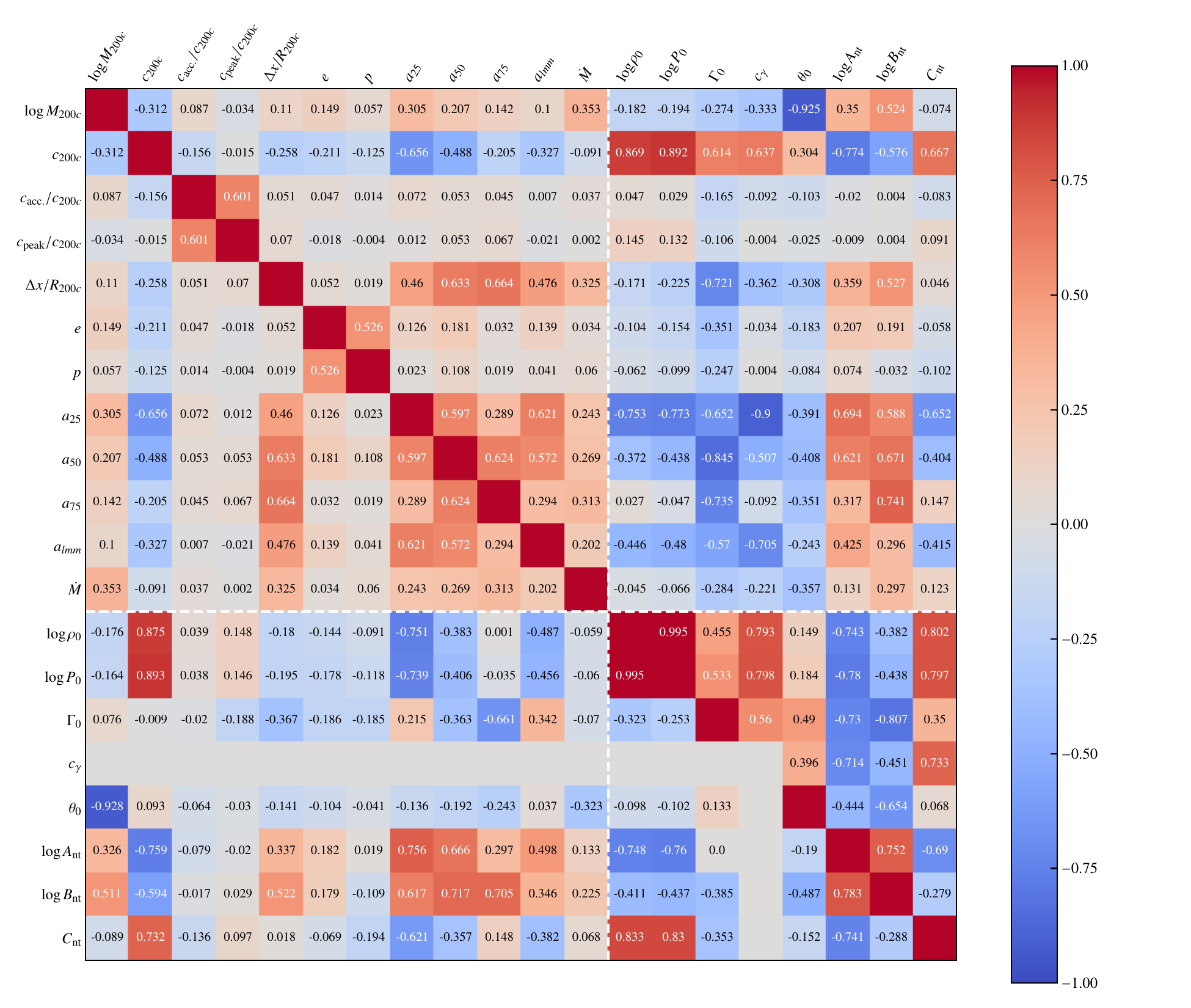}
    \caption{
        Pearson correlation coefficients between components in $\thetahalo$ and $\thetagas$ for the baseline (lower triangle; \S\ref{sec:baseline}) and NR + $\Gamma(r)$ (upper triangle; \S\ref{sec:beyond:Gamma_r}) models.
        The dashed white lines mark the limit between components of $\thetahalo$ and $\thetagas$.
        For the baseline model, $c_\gamma$ is fixed to zero, and therefore does not have correlation with any other parameter.
    }
    \label{fig:params_corrs}
\end{figure*}

In the lower left triangle\footnote{The upper triangle represents the results for another model with varying $c_\gamma$, which will be discussed in \S\ref{sec:beyond:Gamma_r}.} of \cref{fig:params_corrs}, we show the correlation matrices between the components of the concatenation of the $\thetahalo$ and $\thetagas$ vectors, computed for the testing set.
We can identify three blocks:
\begin{itemize}[leftmargin=*]
    \item  The upper left triangular block displays the correlations between the different components of $\thetahalo$.
        We can verify that the components chosen for the input vector do not show any strong colinearity, as no two properties have a correlation greater than 2/3, and only seven pairs from the 12-dimensional space have correlations larger than 1/2.
        These correlations can be further investigated to link halo properties and assembly history, as performed by \eg \citet{2021MNRAS.500.1029L}.
    \item The lower right block covers the correlations between the different components of $\thetagas$.
        We see that $\rho_0$ and $P_0$ are highly degenerate, highlighting the tight relation between gas density and pressure in non-radiative hydrodynamic simulations.
        We note that the two\footnote{Note that we fix $c_\gamma = 0$ in the baseline model.} parameters governing the overall shapes of the density and pressure, $\Gamma_0$ and $\theta_0$, have low correlation.
        Finally, we note that the parameters determining the non-thermal pressure fraction are strongly correlated to each other and to the other parameters of the model.
    \item The lower left rectangular block shows the correlations between the components of $\thetahalo$ and $\thetagas$. 
        We see that $\theta_0$ is strongly $(\rho = 0.93)$ correlated with halo mass.
        On the other hand, $\rho_0$, $P_0$, and the parameters governing the non-thermal pressure fraction are more correlated to halo concentration and early mass assembly history $(a_{25}, \, a_{50})$, while $\Gamma_0$ is more closely related to late $(a_{75}, \, a_{\rm lmm})$ assembly history.
\end{itemize}

\section{Beyond the baseline model} \label{sec:beyond_baseline}

\subsection{Motivation}

In \S\ref{sec:baseline} and \S\ref{sec:baseline:results}, we described the baseline model, and saw that it could provide accurate and precise predictions of ICM thermodynamics.
Nevertheless, while attractive, the baseline model comes with two main drawbacks.

First, the input parameter vector $\thetahalo$ required to make predictions includes a large number of halo properties (listed in \cref{tab:params_thetahalo}).
Not all of these properties are always computed for gravity-only simulations, and some require the computationally expensive reconstruction of halo merger trees.
In addition, all data products of a simulation are not always available to the entire scientific community.
Therefore, maximizing the usability of the \picasso model requires investigating other, more accessible formulations of $\thetahalo$, based on different combinations of halo properties.

Moreover, the baseline model is trained to reproduce gas properties in non-radiative hydrodynamic simulations.
These simulations do not include any modeling of sub-resolution physics, and therefore do not offer a completely representative view of the physical properties of the ICM in the Universe.

In this section, we seek to tackle these challenges by retraining the \picasso model using different training data.
The trained models will be made available along with the baseline model (see \S\ref{sec:code} for more information).

\subsection{Compact model} \label{sec:beyond:compact}

\begin{figure*}[t]
    \centering
    \includegraphics[page=2, width=0.99\linewidth]{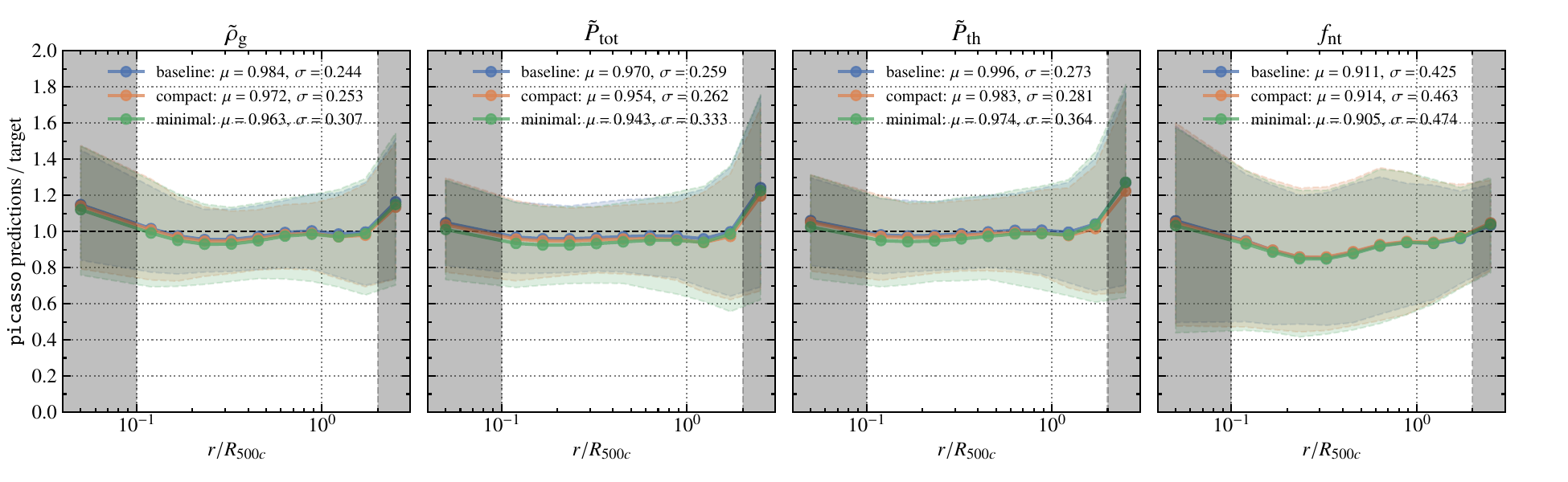}
    \caption{
        Same as \cref{fig:perf:reldiffs_baseline}, showing results for the baseline (blue, \S\ref{sec:baseline}), compact (orange, \S\ref{sec:beyond:compact}), and minimal (green, \S\ref{sec:beyond:minimal}) models.
        We see that reducing the amount of information in the input vector used to predict gas properties results in less precise predictions and a slightly degraded accuracy.
    }
    \label{fig:perf:reldiffs_min_comp}
\end{figure*}

Our first retraining of the \picasso model consists of repeating the baseline analysis while removing information related to the halo mass assembly history from the input vector $\thetahalo$.
This makes every component of $\thetahalo$ accessible from a halo catalog, allowing us to use the model to make predictions without having to access halo merger trees.
The third column of \cref{tab:params_thetahalo} summarizes which properties are included in this model, which we hereafter refer to as the ``compact'' model.

The training procedure is identical to that of the baseline model and uses the same dataset, including the same training/validation/testing splits.
The fully-connected neural network architecture described in \S\ref{sec:baseline:nn} is kept identical, with the exception of the input layer, which is made narrower to reflect the smaller number of input features.
The loss function remains unchanged (eq.~\ref{eq:loss}).

We present the ratios between \picasso predictions and target data in the top row of \cref{fig:perf:reldiffs_min_comp} (orange curves).
Results for the baseline model are also shown for reference (blue curves).
We see that the prediction accuracy is very similar to the baseline results described in \S\ref{sec:baseline:results}.
The precision, however, is slightly degraded, albeit with an average fractional (absolute) increase in scatter of $\simeq 4\% \; (\simeq 1\%)$.

\subsection{Minimal model} \label{sec:beyond:minimal}

Following the reasoning leading to the introduction of the compact model, we further reduce the number of components of the $\thetahalo$ input vector to its minimum.
In this new model---hereafter referred to as the ``minimal'' model---the input vector only includes halo mass $M_{200c}$ and concentration $c_{200c}$ (see \cref{tab:params_thetahalo}).
With such a model, predictions require very minimal information on the halos, making it our most flexible and broadly usable model.
The training process remains unchanged from the compact model (\S\ref{sec:beyond:compact}), except again for the size of the input layer of the fully-connected neural network, which is reduced to two.

The results of the predictions of the minimal model are also presented in the top row of \cref{fig:perf:reldiffs_min_comp} (green curves).
In comparison to the baseline results (blue curves), we see that the accuracy is only slightly degraded, with an average increase in prediction bias of $\simeq 2\%$.
In contrast, the precision is more strongly affected by the removal of information, with a fractional (absolute) increase in scatter of $\simeq 25\% \; (7\%)$, reaching up to $\simeq 33\% \; (9\%)$ for the thermal pressure.
This highlights not only the importance of halo information beyond mass and concentration to predict gas properties, but also the ability of the \picasso model to take advantage of this information when available to make informed, precise predictions.

\subsection{Extension to full-physics hydrodynamics} \label{sec:beyond:fullhydro}

\begin{figure*}[t]
    \centering
    \includegraphics[page=6, width=0.99\linewidth]{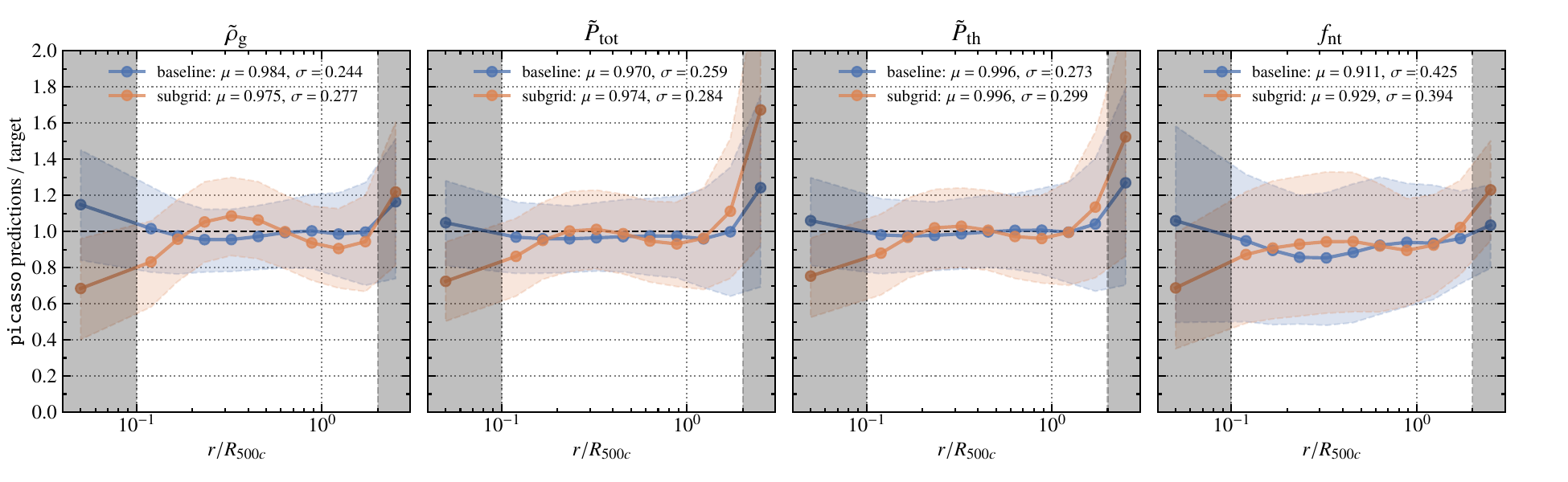}
    \caption{
        Same as \cref{fig:perf:reldiffs_baseline}, showing results for the baseline (blue, \S\ref{sec:baseline}) and subgrid (orange, \S\ref{sec:beyond:fullhydro}) models.
        We see that the model is not able to deliver the same performance for full-physics hydrodynamics as it does for non-radiative simulations, but that the predictions are still quite robust on thermal pressure at intermediate scales $(r/R_{500c} \in [0.15, 1.5])$.
    }
    \label{fig:perf:reldiffs_subgrid}
\end{figure*}

To maximize the practical value of the \picasso model, we seek to evaluate its ability to predict ICM thermodynamic properties that more closely reflect realistic halos, \ie beyond those typically encountered in non-radiative hydrodynamics simulations.
To do so, we re-train the baseline model to reproduce gas properties in our full-physics simulation.
In this new model, which we name the ``subgrid'' model, the input vector $\thetahalo$ remains unchanged from the baseline model, as does the neural network architecture.
The loss function, however, is changed to replace the target data (represented by the non-radiative hydrodynamics thermodynamic profiles in the baseline model) with the full-physics component; \cref{eq:loss} then becomes:
\begin{equation}
    L(\thetanet) = \frac{1}{4} \sum_{i=1}^{4} \frac{1}{N_{\rm bins}}\sum_{j=1}^{N_{\rm bins}} \frac{1}{N_{\rm halos}} \sum_{k=1}^{N_{\rm halos}} \left| \frac{Y_{i, j, k}^{\rm pred} - Y_{i, j, k}^\SG}{Y_{i, j, k}^\SG} \right|.
    \label{eq:loss_sg}
\end{equation}
The optimization process is unchanged from \S\ref{sec:baseline:loss}.

The results of the predictions are shown in \cref{fig:perf:reldiffs_subgrid}.
We see that, averaged in the radial range of interest, the predictions are slightly less biased than for the baseline model; however, the bias is more radius-dependent.
This is particularly noticeable on the gas density, on average biased low by $\simeq 18\%$ for $r/R_{500c} \in [0.1, \, 0.14]$ and high by $\sim 10\%$ for $r/R_{500c} \in [0.27, \, 0.38]$.
We note that this behavior is much less pronounced for the pressures, for which the most biased regions are the first ($r/R_{500c} \in [0.1, \, 0.14]$; $\simeq -9\%$) and last ($r/R_{500c} \in [1.44, \, 2]$; $\simeq +12\%$) radial bins of interest. 
The average scatter in the region of interest is fractionally increased by $\simeq 13\%$ for the pressure and $\simeq 10\%$ for the density, and decreased by $\simeq 7\%$ for the non-thermal pressure fraction (corresponding to absolute changes in scatter of $+3\%, \; +2.5\%, \; -3\%$, respectively).
These results show that the \picasso model demonstrates impressive accuracy in predicting gas properties for full-physics hydrodynamic simulations (with thermal pressure biased by less than a few percent in the radial range $r/R_{500c} \in [0.15, 1.5]$), although it is not quite flexible enough to match the precision and accuracy achieved when training to reproduce non-radiative simulations.

\subsection{Radius-dependent polytropic index} \label{sec:beyond:Gamma_r}

As a final extension to the model, we re-train the baseline (\S\ref{sec:baseline}) and subgrid (\S\ref{sec:beyond:fullhydro}) models while allowing the $c_\gamma$ parameter to vary.
This is motivated by the results presented in \S\ref{sec:beyond:fullhydro}, showing that the model with a fixed $c_\gamma = 0$ was not flexible enough to yield unbiased predictions of the gas properties of halos in a full-physics hydrodynamic simulation.
In particular, the fact that the average prediction bias varies with radius may be an indication of a lack of flexibility in the shape of the model, which releasing the constraint on $c_\gamma$ could address (see fourth column of \S\ref{fig:impact_model_params}).

\begin{figure*}[t]
    \centering
    \includegraphics[page=8, width=0.99\linewidth]{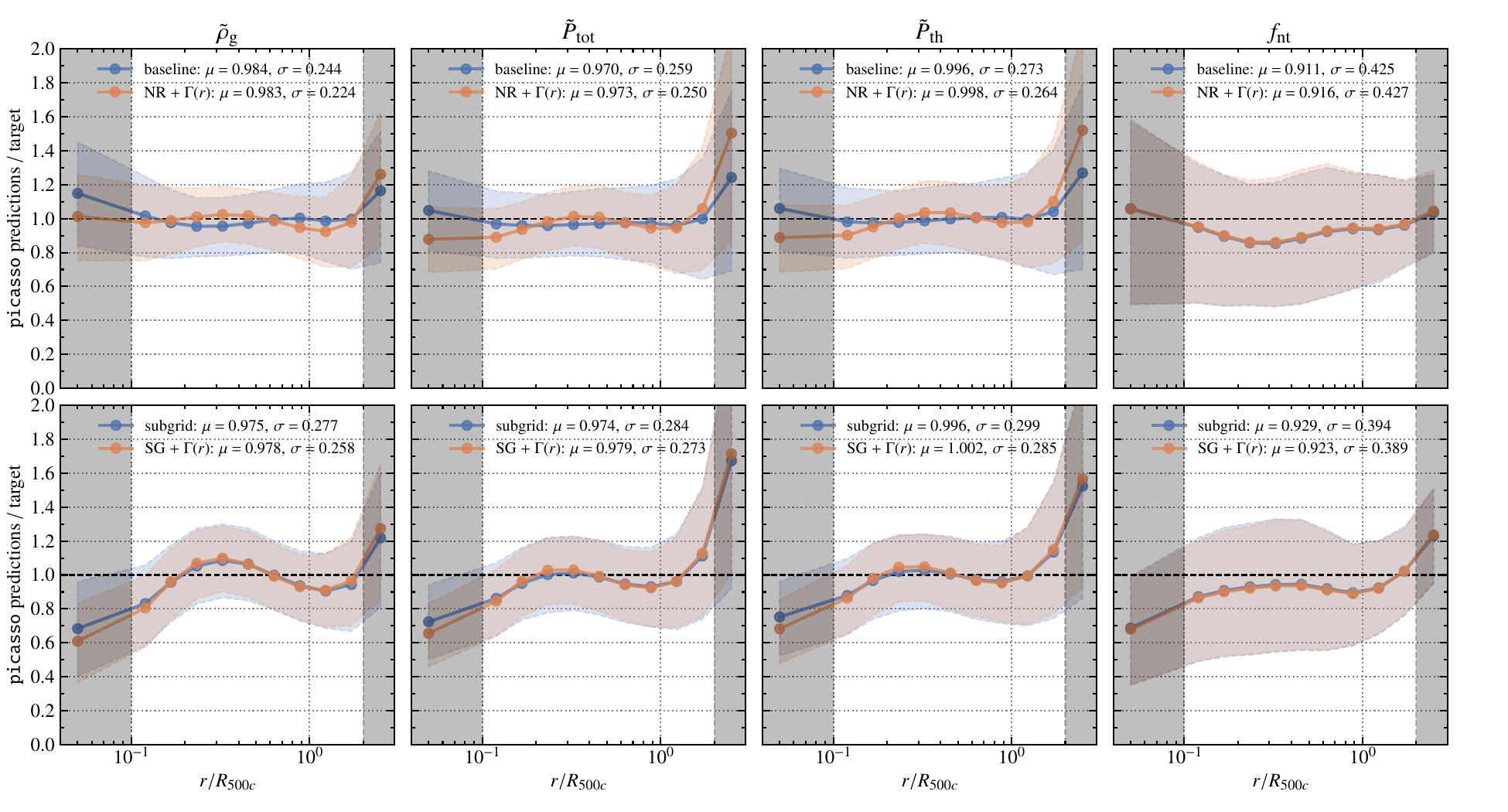}
    \caption{
        Same as \cref{fig:perf:reldiffs_baseline}, showing results for the models with radius-dependent polytropic index (\S\ref{sec:beyond:Gamma_r}).
        \textit{Top row}: baseline (blue) and NR + $\Gamma(r)$ (orange, \S\ref{sec:beyond:Gamma_r}) models.
        \textit{Bottom row}: subgrid (blue, \S\ref{sec:beyond:fullhydro}) and SG + $\Gamma(r)$ (orange, \S\ref{sec:beyond:Gamma_r}) models.
        We see that releasing the $c_\gamma = 0$ constraint has a very small impact on model performance; in particular, it does not attenuate the radial variation of the average prediction bias.
    }
    \label{fig:perf:reldiffs_gamma_r}
\end{figure*}

\paragraph{Non-radiative + $\Gamma(r)$}
We start by re-training the baseline model, using the same training process as described in \S\ref{sec:baseline}, allowing $c_\gamma$ to vary in the $(-1,\, 1)$ range.
The input vector and target data remain unchanged.

Results are shown in the top row of \cref{fig:perf:reldiffs_gamma_r}, and compared to the baseline results.
First, we note that the results on the non-thermal pressure fraction remain largely unchanged, as $c_\gamma$ has no direct impact on $\fnt$, and focus on the predictions for gas density and pressure.
We see that the precision is slightly improved, with relative (absolute) decreases in scatter on density and pressure of $\simeq 8\% \; (2\%)$ and $\simeq 4\% \; (1\%)$, respectively.
We also observe a slight improvement in the accuracy of the predictions, with the average bias on the radial range of interest being reduced for gas density and pressure.
We note that with this new parametrization, the bias on density and pressure exhibits an increased variability with radius, similar in shape to that seen in the subgrid model (\S\ref{sec:beyond:fullhydro}), but with a smaller amplitude.

The resulting correlations between the components of $\thetahalo$ and $\thetagas$ are shown in the upper triangle of \cref{fig:params_corrs}.
We see that $c_\gamma$ is correlated to the other model parameters, in particular to $\rho_0$, $P_0$, $\Cnt$ and $\Ant$.
Moreover, we can see that it most tightly correlates with early mass assembly history ($a_{25}, \, a_{50}$) and merger history $a_{\rm lmm}$.

\paragraph{Subgrid + $\Gamma(r)$}
Finally, we re-train the subgrid model without fixing $c_\gamma = 0$.
The training process is unchanged from \S\ref{sec:beyond:fullhydro}.
We show the results in the bottom row of \cref{fig:perf:reldiffs_gamma_r}, along with those of the subgrid model for comparison.
Again, we see no change in the reconstruction of the non-thermal pressure fraction.
For the density and pressure, we observe a minor improvement in precision (by $\simeq 5\%$ and $\simeq 2\%$, respectively).
As for the accuracy, we can see that it is only marginally improved, and that the radial dependence of the bias is not attenuated by this new formulation.
This points towards a radius-dependent parametrization of the gas polytropic index not being sufficient to make the \picasso gas model provide unbiased predictions of gas thermodynamics in full-physics simulations at all radii.

\section{Numerical implementation} \label{sec:code}

Along with the mathematical description of the model presented in this work (\S\ref{sec:model}, \ref{sec:baseline}) and the assessment of the performances of different trainings of the model (\S\ref{sec:baseline:results}, \ref{sec:beyond_baseline}), we release the \picasso model as a Python package, including numerical implementations of the analytical model described in \refmodeleqs and the different trained models introduced above.
This section describes the numerical implementation of the functions, and the available products and documentation.

\subsection{The analytical gas model} \label{sec:code:analytical}

The \picasso Python package includes functions that can be used to predict gas properties from a gravitational potential distribution and a set of gas model parameters $\thetagas$.
Specifically, the \texttt{picasso.polytrop} module includes functions corresponding to the polytropic gas model of eqs.(\ref{eq:model_rho_P}-\ref{eq:Gamma_r}).
The \texttt{picasso.nonthermal} module provides the formulation of the non-thermal pressure fraction introduced in \cref{eq:model_fnt}.
All of these functions are implemented using \jax \citep{jax2018github}, allowing them to benefit from just-in-time compilation and GPU or TPU acceleration, and to be differentiable with respect to their inputs.

\subsection{Trainable and trained predictors} \label{sec:code:predictors}

The fully-connected neural networks used in the \picasso model are implemented as \texttt{flax.linen} modules, using the \texttt{flax} library, which provides an interface to easily create neural networks using \jax.
In particular, the \texttt{picasso.predictors} module includes the \texttt{PicassoPredictor} class, which uses \texttt{flax.linen} modules to make predictions associated to the gas model.
Specifically, \texttt{PicassoPredictor} objects include two main predicting methods:
\begin{itemize}[leftmargin=*]
    \item \texttt{predict\_model\_parameters} takes as input a set of network parameters $\thetanet$ and a $\thetahalo$ vector, scales it between 0 and 1 (\S\ref{sec:baseline:nn}), and calls the underlying \texttt{flax.linen} module to predict an output vector, which is scaled to produce a predicted $\thetagas$ vector;
    \item \texttt{predict\_gas\_model} takes the same inputs, as well as a gravitational potential distribution, and combines the \texttt{predict\_model\_parameters} function with the \texttt{picasso.polytrop} and \texttt{picasso.nonthermal} modules (\S\ref{sec:code:analytical}) to predict gas thermodynamics associated to the potential distribution.
    It can be used to make predictions for any potential sampling, \eg radial profiles (as used in the training stages presented in this study) or particle distributions.
\end{itemize}
Both functions are also implemented in \jax, allowing them to be compiled just-in-time, hardware-accelerated, and differentiable with respect to their inputs.
In particular, they can be used to write a loss function, taking as input a set of parameters $\thetanet$ for the predicting \texttt{flax.linen} module, and be easily differentiated with respect to the components of $\thetanet$ using \jax.
The loss function and its gradients can then be optimized using efficient gradient descent, for example using \texttt{optax} optimizers \citep{deepmind2020jax}.

The \texttt{picasso.predictors} module also provides the \texttt{PicassoTrainedPredictor} class, inheriting from \texttt{PicassoPredictor}, but using a fixed set of network parameters $\thetanet$.
More notably, \texttt{picasso.predictors} includes \texttt{PicassoTrainedPredictor} instances corresponding to the six trained models presented in this paper, detailed in \cref{tab:trained_models}.

\begin{table}[t]
    \centering
    \begin{tabular}{l c c}
        \toprule
        Object name & Model name & Reference \\
        \midrule
        \midrule
        \texttt{baseline\_576} & baseline & \S\ref{sec:baseline} \\
        \texttt{compact\_576}  & compact  & \S\ref{sec:beyond:compact} \\
        \texttt{minimal\_576}  & minimal  & \S\ref{sec:beyond:minimal} \\
        \texttt{subgrid\_576}  & subgrid  & \S\ref{sec:beyond:fullhydro} \\
        \texttt{nonradiative\_Gamma\_r\_576} & NR + $\Gamma(r)$ & \S\ref{sec:beyond:Gamma_r} \\
        \texttt{subgrid\_Gamma\_r\_576}      & SG + $\Gamma(r)$ & \S\ref{sec:beyond:Gamma_r} \\
        \bottomrule     
    \end{tabular}
    \caption{\normalfont
        List of trained models included in the \texttt{picasso.predictors} module and reference to their description.
        Each object is a \texttt{PicassoTrainedPredictor} object, which includes trained network parameters, and can be called to predict gas model parameters or gas properties.
    }
    \label{tab:trained_models}
\end{table}

\subsection{Performance assessment} \label{sec:code:benchmark}

One of the key features of the numerical implementation of \picasso is its performance.
By using \jax, predictions can easily be compiled just-in-time and take advantage of available hardware (\eg, GPUs) to accelerate computations.
Moreover, in comparison with methods such as baryon pasting, the use of a neural network to predict gas model parameters, instead of relying on the numerical solving of coupled systems of equations, significantly increases the numerical simplicity of the predictions.

To assess the corresponding gain in performance, we benchmark the predictions of both a baryon pasting algorithm and \picasso on a similar problem.
For baryon pasting, we use the implementation described in \bpbc (\S3.2).
For a given halo, we compute the gravity-only matter distribution on a cubic grid with $61^3$ cells, and measure the time needed to solve for conservation of energy and surface pressure when allowing the gas to rearrange from tracing dark matter to following a polytropic equation of state.
For \picasso, we use the pre-trained minimal model to predict gas thermodynamics---specifically the four properties of interest in this work, \ie gas density, total pressure, thermal pressure, and non-thermal pressure fraction---also on a $61^3$ cubic grid, using the compiled \texttt{predict\_gas\_model} function described in \S\ref{sec:code:predictors}.

Both tests are run on the same system, an HP Z8 G4 workstation with two Intel Xeon processors (16 cores each, 2.67 GHz), and an NVIDIA V100 GPU (32GB).
In each case, we run the predictions five times for four different halos (\ie 20 total predictions) and measure the average execution time.
Our implementation of baryon pasting takes an average of $711 \; {\rm ms}$ to make predictions for one halo.
For \picasso, we report an average prediction time of $61 \; {\rm \mu s}$, over four orders of magnitude faster.
Ignoring differences in parallelization schemes for CPU and GPU computations, and neglecting pre-processing overhead, we can extrapolate that the time needed to process $5 \times 10^5$ halos---\ie, roughly the amount of cluster-scale halos expected at $z=0$ in a $\simeq (3 \, h^{-1} {\rm Gpc})^3$ volume, see \eg,\citet{2021ApJS..252...19H}---at this resolution would be about 30 seconds for \picasso, as opposed to 10 hours for our CPU-only implementation of baryon pasting.

We emphasize that this speed-up is the result of many differences between the codes and the models.
First, \picasso benefits from GPU acceleration and just-in-time compilation offered by \jax, in contrast with the CPU-only predictions for our implementation of baryon pasting, and from the use of a (compiled and GPU-accelerated) neural network, as opposed to the solving of coupled systems of equations.
Moreover, it is important to highlight the fact that \picasso predictions rely on a pre-optimized model, for which training represents an overhead that is not reflected in the benchmark; although we do note that most of the time associated to the training corresponds to the data preparation and pre-processing described in \S\ref{sec:data:576} (\ie reading in large simulation data products, matching halos across simulation runs, and measuring thermodynamic profiles from particle data); the training itself (\S\ref{sec:baseline:loss}) is completed in less than a minute on the system described above.

\subsection{Availability}

The code is publicly available on Github\footnote{\url{https://github.com/fkeruzore/picasso}}.
We also provide an online documentation\footnote{\url{https://picasso-cosmo.readthedocs.io}}, which includes instructions to install the package, a documentation of the different modules and functions, and several example notebooks.

\section{Conclusions and perspectives} \label{sec:end}

We have introduced \picasso, a model allowing the prediction of the thermodynamic of intracluster gas from the properties of dark matter halos in gravity-only simulations.
The model, described in \S\ref{sec:model}, combines an analytical mapping between the gravitational potential distribution in a halo and ICM thermodynamics with a machine learning model predicting the parameters of this gas model from halo properties.
This combination presents three main advantages:
\begin{itemize}[leftmargin=*]
    \item \textit{Numerical efficiency}: neural networks are straightforward mathematical objects, and can make predictions more efficiently than purely analytical models based on solving coupled equations corresponding to a physical transformation (\eg, \bpbc and similar works);
    \item \textit{Speed and differentiability}: The use of \jax for the numerical implementation of the \picasso model enables hardware acceleration and makes predictions differentiable, enabling efficient gradient-based model optimization, and opening the possibility of high accuracy gas models in simulation-based inference \citep[\eg,][]{2020PNAS..11730055C, 2024MNRAS.527.1244S, 2024arXiv240710877L};
    \item \textit{Flexibility}: Because the gas model predicts gas properties from gravitational potential distributions, a trained model can be used to make predictions from a variety of inputs.
    One may use \picasso with inputs ranging from a simple halo catalog (\eg assuming a spherically-symmetric potential model), to the full particle output of an $N-$body simulation, taking full advantage of these large data products to capture the three-dimensional shape of halos.
\end{itemize}

We train the \picasso model by forward modeling gas properties from gravity-only halos, and training the model to maximize the agreement between these predictions and the thermodynamic properties of the halos found in matching hydrodynamic simulations (\S\ref{sec:data}, \ref{sec:baseline}).
We perform several alternative trainings (\S\ref{sec:beyond_baseline}), corresponding to different inputs availability and simulation physics, reaching the following conclusions:
\begin{itemize}[leftmargin=*]
    \item With access to a large variety of halo properties to make predictions from, the \picasso model can be trained to reproduce the ICM thermodynamics of halos in a non-radiative hydrodynamic simulation with remarkable accuracy and precision, with few-percent-level bias and 15 to 20\% scatter on gas density and pressure across a radial range \fullrrange\ (\S\ref{sec:baseline:results}).
    \item When trained to make predictions from a smaller subset of halo properties, the accuracy and precision are slightly altered, but remain promising.
    In particular, from a very minimal input consisting of only halo mass and concentration, bias only increases up to $<3\%$, while scatter is fractionally increased by $\sim 30\%$ (\S\ref{sec:beyond:minimal}).
    \item The \picasso model can be trained on full-physics hydrodynamic simulations, although the predictions become less accurate, with an average bias showing variation with distance from the halo center, ranging between $-9\%$ and $+10\%$ on the radial range \fullrrange.
    The precision is also slightly degraded, with a fractional increase in scatter of around 10\% (\S\ref{sec:beyond:fullhydro}).
    \item Generalizing the model to a radius-dependent gas polytropic index $\Gamma$ does not improve the accuracy of the predictions, and only marginaly decreases their scatter (\S\ref{sec:beyond:Gamma_r}).
\end{itemize}

The \picasso model is made available to the community as a Python package, including the analytical gas model and the trained predictors presented in this work, and an extensive online documentation, including running examples.
Given its high accuracy, precision, and flexibility, \picasso can be used to produce high-quality synthetic datasets, which we believe will prove a useful capability for the cluster cosmology community.

\subsection{Future work} \label{sec:end:future}

This article, presenting the model and its performance, will be the first in a series of publications dedicated to \picasso.
The next entry in the series will focus on creating tSZ maps using the trained models and gravity-only simulation presented above, and validating them against those created from the matching hydrodynamic volumes \citep{keruzore_576_maps}.
Future work will include the exploitation of the model to create sky maps of the tSZ effect, in particular from highly-used gravity-only simulations evolved using the \HACC solver \citep{habib_hacc_2016} such as the OuterRim \citep{2019ApJS..245...16H}, LastJourney \citep{2021ApJS..252...19H} and New Worlds \citep{2024arXiv240607276H} simulations, and the validation of these maps against observations of the tSZ effect from, \eg, the South Pole Telescope \citep{2022ApJS..258...36B}.

Furthermore, while we have seen that the \picasso gas model is efficient at predicting intracluster gas properties from gravity-only simulations, it can still be improved further.
Several specific aspects will be investigated, including:
\begin{itemize}[leftmargin=*]
    \item Improving the gas model: we have shown that there is room for improvement in the ability of the \picasso model to predict gas properties from realistic simulations with sub-resolution physics.
    Such improvements will likely require adapting the analytical model of \refmodeleqs to include more subtle effects, in particular related to radiative cooling, star formation, and feedback from active galactic nuclei;
    
    \item The impact of cosmological parameters and baryonic physics: by re-training \picasso on hydrodynamic simulations with varying underlying cosmologies and sub-resolution prescriptions.
        Thanks to the advent of exascale systems such as Frontier and Aurora, such suites of simulations, spanning large volumes in the cosmological and subgrid parameter spaces, will be run with volumes large enough to provide enough cluster-scale objects to re-train \picasso models.
        These trained models will be used to investigate the difference between predicted cluster gas properties with simulation parameters, as well as to build emulators, enabling the emulation of hydrodynamic simulations for arbitrary sets of cosmological and subgrid model parameters;
    
    \item The impact of redshift: as presented here, \picasso was trained on halos at $z=0$.
        While the trained models can still be used to make predictions at higher redshifts---since they predict redshift-evolving gas properties through \cref{eq:model_rho_P}---this ignores redshift evolution beyond self-similarity.
        Training the model on different redshift snapshots---and interpolating between training redshifts when using the model to make predictions from lightcone data---will be an important extension to ensure the accuracy of the sky maps.
        A first assessment of the accuracy and precision of the baseline model, trained at $z=0$, when making predictions at higher redshifts, is presented in Appendix \ref{sec:app:z0.5}.
        
    \item Training beyond profiles: we have focused on learning the mapping between potential and gas properties on azimuthal profiles.
        While this does not impact the ability to use the learned model to make predictions for arbitrary potential distributions, it implies that some of the information on halos was lost during the training phase.
        In principle, one can generalize the training method presented in \S\ref{sec:baseline:loss} to train the model while retaining multi-dimensional information, for example by replacing the radial profiles of halo properties with three-dimensional grids.
        This comes at a large computational cost, as it significantly increases the memory required during training.
        Nonetheless, owing to the scalability of \jax and \texttt{flax}, this can be circumvented by training in batches on high-performance supercomputers, in particular by taking advantage of large GPU servers.
        
    \item Probabilistic predictions: the use of simple fully-connected neural networks to predict gas model parameters restrains us to making point-like predictions.
        In order to efficiently create large numbers of mock datasets, the ability to predict probability distributions in the parameter space can be useful, as one can then randomly sample said distributions and create several datasets from one gravity-only simulation, and propagate model uncertainty to analyses.
\end{itemize}

\section*{Acknowledgements}

\small
Argonne National Laboratory's work was supported by the U.S. Department of Energy, Office of Science, Office of High Energy Physics, under contract DE-AC02-06CH11357, and used resources of the Argonne Leadership Computing Facility, which is a DOE Office of Science User Facility.
This research also used resources of the Oak Ridge Leadership Computing Facility, which is a DOE Office of Science User Facility supported under Contract DE-AC05-00OR22725.
We gratefully acknowledge the computing resources provided on Improv, a high-performance computing cluster operated by the Laboratory Computing Resource Center at Argonne National Laboratory.
This work was initiated at the Aspen Center for Physics, which is supported by National Science Foundation grant PHY-2210452.

\paragraph{Software}
\picasso relies on libraries from the \jax ecosystem, in particular on \jax \citep[for GPU-compatible and differentiable numerical computations;][]{jax2018github}, \texttt{flax} \citep[for neural network-based models;][]{flax2020github} and \texttt{optax} \citep[for model optimization;][]{deepmind2020jax}.
The cosmological simulations presented in this work were evolved using the (\CRK-)\HACC solver \citep{habib_hacc_2016, frontiere_simulating_2023}.
This research made use of various Python libraries, including \texttt{astropy} \citep{astropy_collaboration_astropy_2018}, \texttt{mpi4py} \citep{dalcin_mpi4py_2021}, \texttt{numpy} \citep{harris2020array}, and \texttt{scipy} \citep{virtanen_scipy_2020}.
Figures were prepared using \texttt{matplotlib} \citep{hunter_matplotlib:_2007} and \href{https://draw.io}{draw.io}.
\normalsize


\bibliographystyle{aasjournal}
\bibliography{picasso}

\begin{appendix}

\section{Predictions at higher redshift} \label{sec:app:z0.5}

\begin{figure*}
    \centering
    \includegraphics[width=0.99\linewidth]{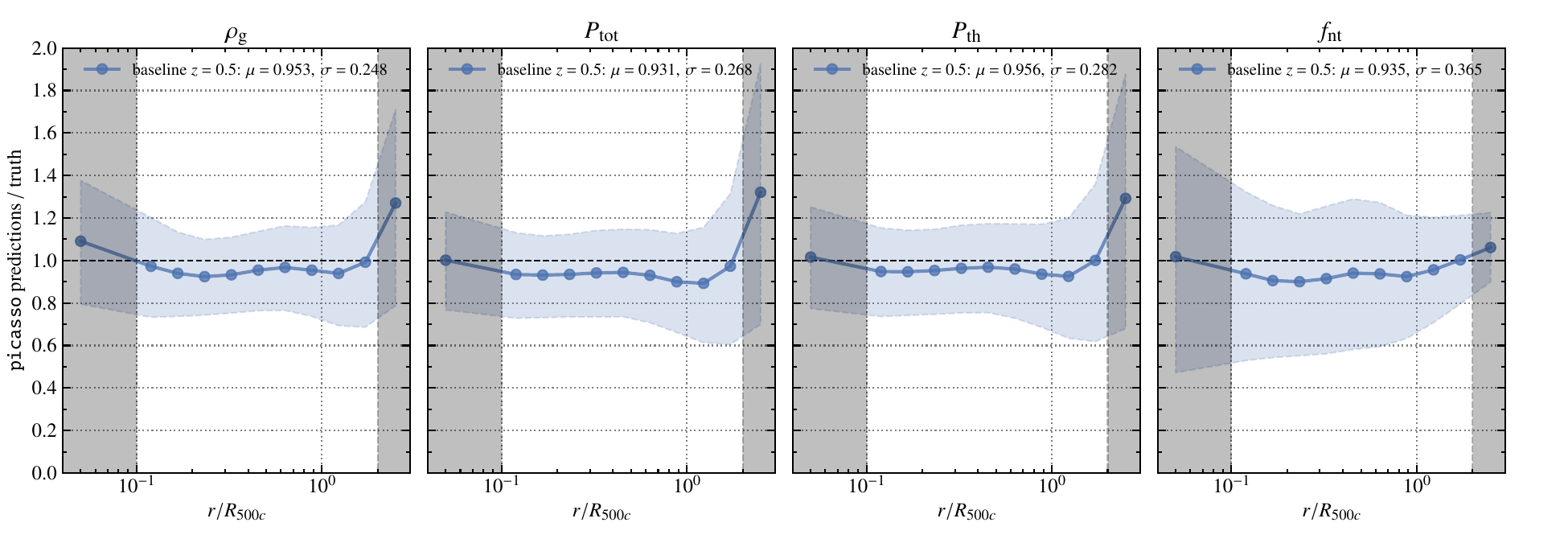}
    \caption{
        Same as \cref{fig:perf:reldiffs_baseline}, showing results for predictions made by the baseline model (trained at $z=0$, \S\ref{sec:baseline}) for halos at $z=0.5$.
        We see that the precision of the predictions is mostly unaffected, while the accuracy is slightly degraded, with the average bias on thermal pressure going from 0.4\% at $z=0$ to 4.4\% at $z=0.5$.
    }
    \label{fig:perf:reldiffs_0.5}
\end{figure*}

In this work, we have only trained the \picasso model at $z=0$.
As discussed in \S\ref{sec:end:future}, while the model does include the possibility of a redshift evolution of the gas thermodynamic properties---since it predicts scaled density, $\rho / \rhocrit$, and pressure, $P / P_{500c}$, as described in \cref{eq:model_rho_P}---this ignores the possibility of a more complex redshift evolution.
Moreover, the non-thermal pressure fraction does not include any explicit redshift evolution.
Therefore, using the \picasso model trained at $z=0$ to infer gas properties at higher redshifts might lead to less robust predictions.

As a first assessment of this potential loss in performance, we use the baseline model, trained at $z=0$, to predict gas properties for halos at higher redshift.
Using the simulations presented in \S\ref{sec:data:576} at $z=0.5$, we follow \S\ref{sec:data:halo_match} in matching halos in the gravity-only and non-radiative hydrodynamic volumes, and \S\ref{sec:baseline:thetahalo} in computing the $\thetahalo$ input vectors.
We then use the procedure presented in \S\ref{sec:data:profiles} to measure the corresponding gravitational potential profiles in the gravity-only volume, and thermodynamic profiles in the non-radiative hydrodynamic one.
Then, the already trained baseline model (\S\ref{sec:baseline}) is used to predict the gas properties from the full set of gravity-only halos, following \S\ref{sec:baseline:fwmod}.
For each halo, we then compute the ratio between the \picasso predictions of gas thermodynamics and the properties measured in the hydrodynamic run.

Results are shown in \cref{fig:perf:reldiffs_0.5}.
Comparing the performances of the predictions with those of the baseline model evaluated at $z=0$ (\cref{fig:perf:reldiffs_baseline}), we see that the precision is mostly unaffected.
However, we note that the predictions are marginally biased low on average, slightly more than at $z=0$ (about 4\% on average for \fullrrange for the thermal pressure).
This is not unexpected, as the parameters of our gas model are known to evolve with redshift \citep[see fig. 4 of][]{2023OJAp....6E..43K}, which is not taken into account in our neural network predictions.
While the resulting bias is quite small, it motivates further investigations of the impact of redshift on training.
Future releases of trained \picasso models will focus on either training on lightcone data---treating redshift as another component of the $\thetahalo$ vector---or on multiple redshift snapshots independently, with an interpolation of the predictions between redshifts.
Nevertheless, this low level of inaccuracy is encouraging, as it shows that little more is needed over the current implementation of the \picasso model to achieve unbiased predictions at any redshift.
\vspace{10pt}

%
%
%

\end{appendix}

\end{document}